\documentclass[aps,prb,10pt,superscriptaddress,amsmath,amssymb,twocolumn,nofootinbib]{revtex4-2}
\usepackage{graphicx,bm,amsmath}
\usepackage[colorlinks]{hyperref}

\usepackage[bbgreekl]{mathbbol}
\usepackage{xcolor}
\usepackage{braket}
\usepackage{verbatim}
\usepackage{float}
\usepackage{appendix}

\usepackage{subcaption}
\usepackage{tikz}
\usetikzlibrary{decorations.pathreplacing,arrows.meta,positioning}

\usepackage{ragged2e}

\begin{document}
\title{Thermalization and quantum chaos in a pure gauge discrete non-Abelian lattice gauge theory} 

\author{Marco Vaia} 
 \affiliation{Dipartimento di Fisica, Universit\`a di Trento,  Via Sommarive 14, I-38123 Trento, Italy}
\author{Edoardo Ballini}
 \affiliation{Pitaevskii BEC Center, CNR-INO and Dipartimento di Fisica, Universit\`a di Trento,  Via Sommarive 14, I-38123 Trento, Italy}
 \affiliation{INFN-TIFPA, Trento Institute for Fundamental Physics and Applications, Via Sommarive 14, I-38123 Povo, Trento, Italy}
 \affiliation{Institute for Quantum Optics and Quantum Information,
Austrian Academy of Sciences, Innsbruck 6020, Austria}
\affiliation{Institute for Theoretical Physics, University of Innsbruck, Innsbruck 6020, Austria}
\author{Alberto Bottarelli}
 \affiliation{Pitaevskii BEC Center, CNR-INO and Dipartimento di Fisica, Universit\`a di Trento,  Via Sommarive 14, I-38123 Trento, Italy}
 \affiliation{INFN-TIFPA, Trento Institute for Fundamental Physics and Applications, Via Sommarive 14, I-38123 Povo, Trento, Italy}
\author{Matteo M. Wauters} 
 \affiliation{Pitaevskii BEC Center, CNR-INO and Dipartimento di Fisica, Universit\`a di Trento,  Via Sommarive 14, I-38123 Trento, Italy}
 \affiliation{INFN-TIFPA, Trento Institute for Fundamental Physics and Applications, Via Sommarive 14, I-38123 Povo, Trento, Italy}
 \author{Philipp Hauke}
  \affiliation{Pitaevskii BEC Center, CNR-INO and Dipartimento di Fisica, Universit\`a di Trento,  Via Sommarive 14, I-38123 Trento, Italy}
 \affiliation{INFN-TIFPA, Trento Institute for Fundamental Physics and Applications, Via Sommarive 14, I-38123 Povo, Trento, Italy}

\begin{abstract}
Gauge theories constitute the basis of our microscopic description of fundamental physical laws, and they are a cornerstone of contemporary scientific research. 
Yet, how such systems thermalize far from equilibrium remains a fundamental open question at the crossroads of statistical, high-energy, and condensed-matter physics. 
In this work, we investigate the thermalization properties of non-Abelian lattice gauge theories, focusing on the discrete gauge group $D_3$, which is ideally suited for current qudit-based quantum simulators. 
Specifically, we study a linear plaquette chain in a gauge-fixed formulation that removes unphysical degrees of freedom. 
A numerical analysis of the Hamiltonian spectrum and expectation values of local operators shows a behavior broadly aligned with the Eigenstate Thermalization Hypothesis (ETH).
We also study thermalization through the real-time dynamics of the entanglement spectrum, confirming the emergence of random-matrix level statistics after an initial transient.
Our results expand the rich phenomenology of dihedral non-Abelian gauge systems. Once the relevant symmetry structure is properly resolved, the model displays the same qualitative ETH-based thermalization scenario observed in Abelian gauge models and generic non-integrable quantum many-body systems.
\end{abstract}

\maketitle

\section{Introduction}
The effective thermalization of local observables in isolated quantum systems, as described by the \textit{Eigenstate Thermalization Hypothesis (ETH)}, is a cornerstone of out-of-equilibrium quantum many-body physics~\cite{Rigol2008, D_Alessio2016, Mori2018,Buca2023}. 
Although quantum mechanics lacks the rigorous chaos hierarchy of classical dynamical systems, the ETH has proved extremely effective in describing the long-time unitary quench dynamics and in linking the emergence of effective thermal ensembles to the non-integrability of the underlying Hamiltonian~\cite{Srednicki1994}.
However, there are notable exceptions that violate ETH or require modified formulations, such as many-body localization~\cite{Geraedts2016, Alet2018}, integrable systems~\cite{Rigol2007,Alba_PRB2015,Vidmar_JSTAT2016}, non-Hermitian and open systems~\cite{Cipolloni_PRB2024,SinghaRoy_PRL2025,Almeida_PRE2026}, and quantum scars~\cite{Bernien2017, Serbyn2021,Moudgalya_2022}. 

In this context, lattice gauge theories (LGTs)~\cite{Rothe2012,Dalmonte2016,Banuls2020,Bauer_PRX2023} provide a fascinating setting to study thermalization~\cite{Zhou2022,Mueller2022,Mueller2025, Berges2018}, as they are the main framework to describe strong coupling in fundamental physics as well as the emergence of low-energy collective phenomena in condensed matter.
The presence of an extensive number of local conserved charges implies that ETH is inevitably broken in the full Hilbert space.
However, when restricting the description to a single superselection sector, the situation is more subtle, and the thermalization properties (or lack thereof) depend on many factors, such as the gauge group, possible truncations to a discrete Hilbert space, or the lattice geometry.
Non-Abelian groups add further puzzles to the picture. Non-Abelian global symmetries modify the standard ETH scenario~\cite{Murthy2023, Majidy2023, Patil_PRB2025}, and in LGTs they induce more complex and non-commuting local constraints on the quantum dynamics~\cite{Zohar2015, Davoudi2021, Halimeh2022, Ballini2025}.
Moreover, understanding thermalization properties in non-Abelian gauge theories could provide precious insights into some outstanding open problems in fundamental physics, such as the early-universe thermalization or the fast thermalization observed in heavy-ion collisions~\cite{Kofman1997,Berges2021, Florio2025}.

Yet, direct calculations of non-Abelian LGTs through the ETH lens are still at a very early stage, partly also because of the considerable computational complexity of performing full diagonalization on such systems, which typically have large local Hilbert spaces. ETH has been characterized in detail in instances of SU(2) LGTs with small truncations to irreducible representations~\cite{Xiaojun2023,Ebner2024, Ebner2025,Das2026}.
However, the small Hilbert space truncation adds further constraints that may affect the thermalization, or its violations, in such systems and is suitable only for studying strong-coupling regimes. For instance, 1+1D SU(2) with dynamical matter is a paradigmatic example for disorder-free localization and Hilbert space fragmentation~\cite{Cataldi_PRL2026,ciavarella_2025}.

In this paper, we approach the topic from a different perspective by considering a discrete non-Abelian gauge symmetry, the dihedral group $D_3$. 
Although it is not directly related to QCD, it is a useful testbed to investigate the physics of non-Abelian LGTs in a simplified and well-controlled setting. 
Discrete groups, including discrete subgroups of continuous ones, can be suitably implemented in qudit quantum-simulation platforms~\cite{Wang2020,Low2020,Ringbauer2022,Chi2022, Fischer2023,Liu2023, Burshtein2026}, since their finite dimension allows for exact encoding onto qudit levels~\cite{Gonzalez_Cuadra2022, Gustafson_PRD2022, Zache2023, Gaz2025, Meth2025}. At the same time, discrete non-Abelian LGTs display rich phenomenology, bridging condensed matter, high-energy physics, and quantum information~\cite{Munk2018,albert2021,Mariani2023,Santra2025,Popov2025,Lo2025,Lo2026}.

\begin{figure}[t]
    \centering
    \includegraphics[width=\linewidth]{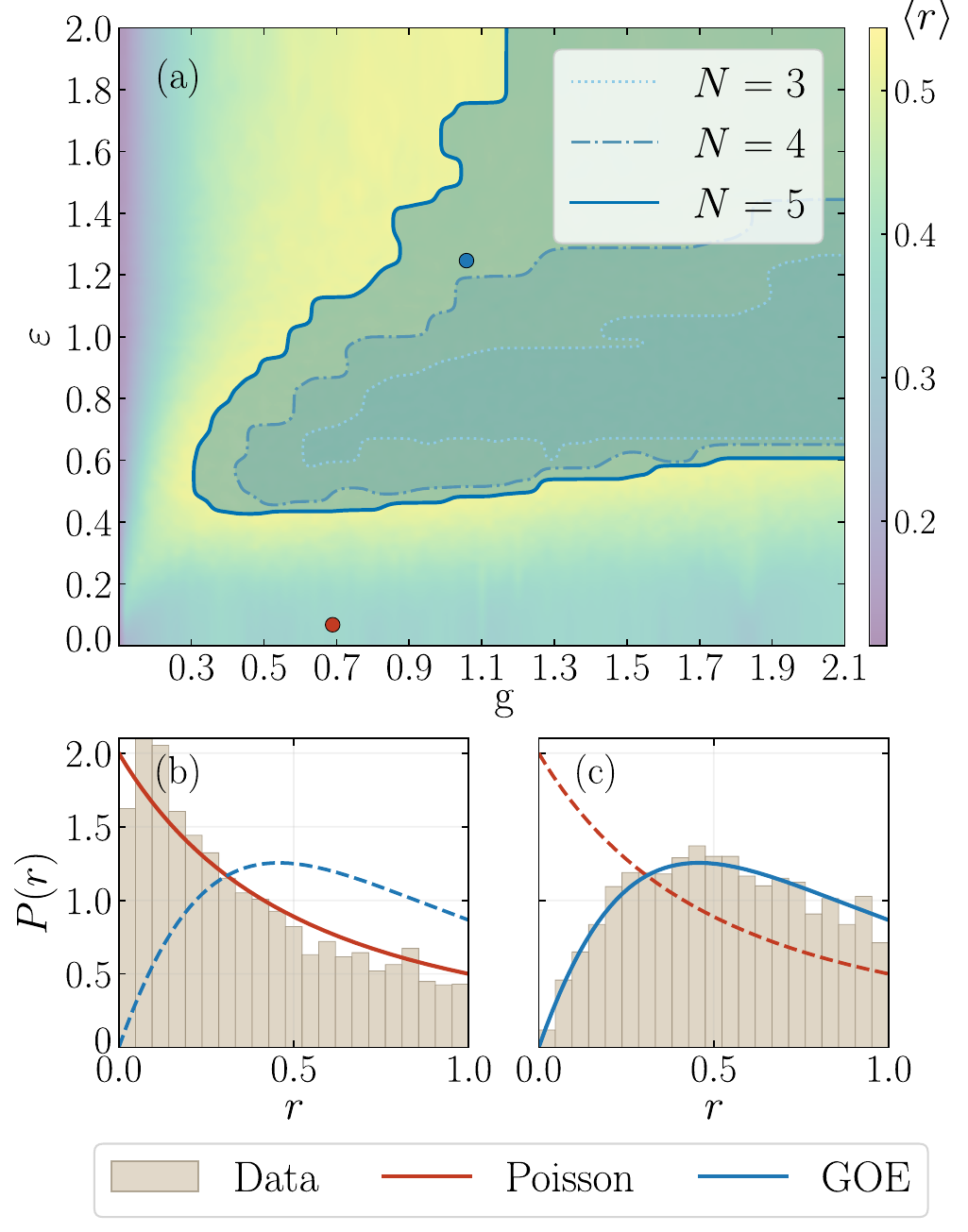}
    \caption{ \justifying
    Overview of the spectral-statistics phase diagram. 
    Panel (a) shows the heatmap of the average minimal gap ratio, $\braket{r}$, in the parameter space spanned by the coupling $\mathrm{g}$ and the plaquette deformation $\varepsilon$, for $N=5$. The shaded regions indicate where the system displays GOE-like spectral statistics within the chosen tolerance $(\pm 4 \%)$, after interpolation and smoothing for visualization, with darker shades corresponding to larger system sizes. Their growth with the system size $N$ suggests that the chaotic regime becomes progressively more robust. Panels (b) and (c) show two representative points for $N=5$: one with accidental Poisson-like level statistics and one in the chaotic regime with GOE statistics.
    The spectrum is restricted to the even sector of the vertical-inversion symmetry.
    }
    \label{fig:spectralstat}
\end{figure}

In our work, we focus on a ladder geometry with a pure gauge Hamiltonian, where we apply a gauge fixing procedure to remove non-physical degrees of freedom.
Inspired by Ref.~\cite{Munk2018}, we also consider deformations of the plaquette operator to break the global $D_3$ symmetry, in a fashion reminiscent of twisted interactions in chiral clock models~\cite{Howes1983,Fendley_2012,Moran_PRB2017,Wauters2025}. This modification provides a tunable parameter to enhance the non-integrable behavior in the system.
We study a number of ETH indicators: the minimal energy gap ratio statistics~\cite{Berry1977,Mehta2004, Bohigas1984, Oganesyan2007, Atas2013}, the matrix elements of local operators in the energy basis~\cite{Deutsch1991,Srednicki1994,Srednicki1999,Rigol2008,D_Alessio2016}, and the entanglement Hamiltonian~\cite{Li2008,Kokail2021,Dalmonte2022,Mueller2022,Mueller2025}. 
Contrary to other lattice gauge theories~\cite{Surace_PRX2020,Banerjee2021,Saptarshi2022, Yao_PRB2022, Halimeh_2023, Calajo2025, Aramthottil2022, Hayata2023_1, Sau2024, Ge2024}, in the parameter regimes analyzed we find no signatures of scar states. 
Despite the small system sizes that we can tackle numerically, a chaotic region dominates the phase diagram spanned by the coupling strength  $\mathrm{g}$ and the deformation of plaquette operators $\varepsilon$, whose size grows for larger systems, as summarized by the behavior of the average gap ratio $\braket{r}$ in Fig.~\ref{fig:spectralstat}.
At small coupling or small symmetry-breaking term, we observe a region where the level statistics deviates significantly from that of a Gaussian orthogonal ensemble (GOE). As the system size increases, this region is pushed closer to the undeformed model at $\varepsilon=0$, suggesting that the chaotic regime becomes progressively more robust in larger systems.

Our findings provide a complementary setting to Lie groups to study thermalization in non-Abelian LGTs, and can help in distinguishing the phenomenology due to a specific group choice from effects associated with general non-Abelian symmetries. 
Furthermore, the fast development of qudit devices raises the compelling perspective of performing experimental investigations of such models in the near future, going beyond the system sizes accessible to classical simulators. 

The rest of the paper is organized as follows. Section~\ref{sec:model} gives a short review of the mathematical structure of non-Abelian LGTs and describes the Kogut--Susskind formulation we adopt in this paper, along with the gauge fixing procedure.
In Sec.~\ref{sec:ETH}, we characterize the spectral statistics and thermalization properties of the $D_3$ ladder, while the entanglement properties of the eigenstates are discussed in Sec.~\ref{sec:entanglement}.
We summarize our results and discuss future perspectives in Sec.~\ref{sec:conclusions}.
Further details are provided in the Appendices. The gauge fixing for a non-Abelian group is described in Appendix~\ref{app:Ham}. 
Further numerical results characterizing the ETH on the $D_3$ ladder can be found in Appendix~\ref{app:more_res}.

\section{Non-Abelian Gauge Ladder}\label{sec:model}
\begin{figure}[t]
    \centering
    \includegraphics[width=\linewidth]{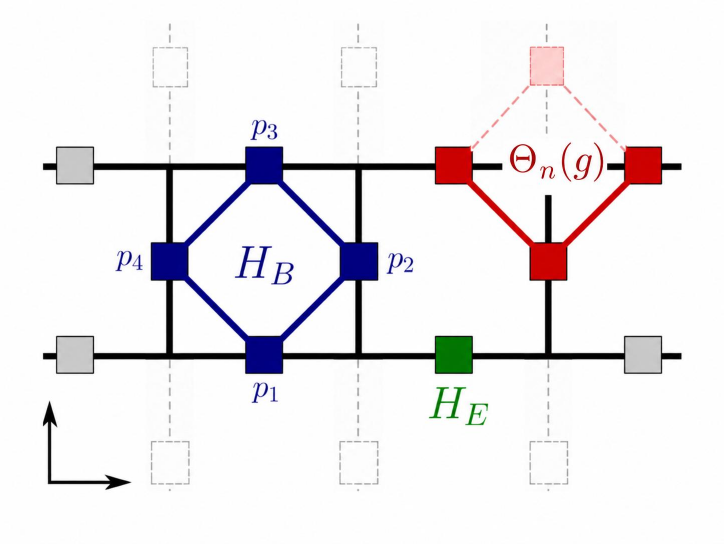}
    \caption{\justifying $D_3$ pure gauge theory on a square lattice with gauge degrees of freedom on the links. In our numerical calculations, we consider a quasi-one-dimensional ladder. 
    Highlighted are a four-body term entering the magnetic Hamiltonian $H_B$, a single-body term of the electric Hamiltonian $H_E$, and a local gauge transformation $\Theta_{n}(g)$. The arrows in the bottom-left corner indicate the directionality of the lattice.
    }
    \label{fig:Latticegen}
\end{figure}

In this work, we focus on a pure gauge theory in $2+1$ space-time dimensions in the Kogut--Susskind Hamiltonian formulation~\cite{Kogut1975}. The gauge group is the discrete non-Abelian dihedral group $D_3$.
We consider a square lattice with gauge variables residing on its links, where, to make the system amenable to numerical analysis, we focus on a ladder geometry, see Fig.~\ref{fig:Latticegen}. 
To provide the necessary background, we first review the basic formalism for the Hamiltonian formulation of non-Abelian LGT models, following the notation of Ref.~\cite{Zohar2015}. 

\subsection{Kogut--Susskind formulation of dihedral pure gauge LGT}

Consider a discrete non-Abelian group with elements $g \in G$ and order $|G| \in \mathbb{N}$. We denote a unitary matrix representation of the group elements as $D^j_{mn}(g)$, where $j$ labels the irreducible representations and $m,n$ are the row and column indices of the matrix. States labeled by the representation indices $|jmn\rangle$ form a complete basis. This representation basis is connected to the group element basis $\ket{g}$ through the great orthogonality theorem~\cite{Serre1977}, which also describes the generalized quantum Fourier transform of the group, 
\begin{equation}
    \bra{g}jmn\rangle = \sqrt{\frac{\dim(j)}{|G|}} D^j_{mn}(g) \ .
\end{equation}
In the following, we will label the computational basis states with the group elements $\ket{g}$. 

The states $\ket{g}$ transform according to the group multiplication. In non-Abelian groups, we must distinguish between left- and right-multiplication operators, which we define in the following way:
\begin{equation}
\begin{aligned}
    \theta^L(h) \ket{g} &= \ket{hg} \ , \\
    \theta^R (h) \ket{g} &= \ket{gh^{-1}} \ . \\
\end{aligned}
\end{equation} 
Finally, the group parallel transporter $U^j$ associates the states $\ket{g}$ with the $j$-th representation of the corresponding group element $U^j_{mn} \ket{g} = D^j_{mn}(g) \ket{g}$.

With these ingredients, we can describe the pure gauge Kogut--Susskind Hamiltonian in $2+1$ dimensions~\cite{Kogut1975}, which reads
\begin{equation}
    H_\mathrm{KS} = H_B+H_E \ .
    \label{eq:HKS}
\end{equation}
The first term of the Hamiltonian describes the energy associated with (non-Abelian) magnetic fluxes. Its simplest formulation is (we work with natural units $c = \hbar = 1$)
\begin{equation}
    H_B = - \frac{1}{\mathrm{g}^2} \sum_{\mathrm{plaquettes}} \Re \left(\mathrm{Tr}\left[U^j_{p_1}U^j_{p_2}U^{ j \dagger}_{p_3} U^{ j \dagger}_{p_4}\right]\right) \ ,
\end{equation}
where $\mathrm{g}^2$ is the coupling and the sum runs over all plaquettes in the lattice. $U_{p_i}^j$ is the parallel transporter in the irreducible representation $j$ associated with link $p_i$ located in the plaquette (see Fig.~\ref{fig:Latticegen}), with  $p_1,\ldots,p_4$ taken in anticlockwise order and starting from the lower edge, see Fig.~\ref{fig:Latticegen}. The directionality of the lattice links determines whether we have to take the operator $U^j_{p_\alpha}$ or its Hermitian conjugate. Hereafter, we omit the superscript $j$, as we always take $U_{p_\alpha}$ in the two-dimensional irrep $\mathbf{2}$ of $D_3$.

The second term of the Hamiltonian in Eq.~\eqref{eq:HKS}, $H_E$, represents the action of a local electric field. It is diagonal in the representation basis. Since gauge invariance does not impose a particular choice of its eigenvalues~\cite{Zohar2015, Mariani2023}, its most general form reads
\begin{equation}
    H_E = \mathrm{g}^2 \sum_{\ell\in \mathrm{ links}} \sum_{j} \alpha_j \Pi_{\ell}^j \ ,
    \label{eq:elecham}
\end{equation}
where $\Pi_{\ell}^j$ is the projector onto the $j$ representation acting on link $\ell$, and $\alpha_j$ is a function of the representation only.

Physical observables are invariant under local gauge transformations $\Theta_{n}(g)$, which act non-trivially on all links connected to a given vertex $n$ of the lattice $\Lambda$. In pure gauge systems, $\Theta_{n}(g)$ applies left group multiplications to the outgoing links and right multiplications to the ingoing ones~\cite{Zohar2015}, 
\begin{equation}
    \Theta_{n} (g) = \prod_{o} \theta^L_{o} (g)\prod_{i} \theta^R_{i}(g) \ .
    \label{eq:gaugetrans}
\end{equation}
Gauge-invariant (physical) states also transform trivially under the symmetry group $\Theta_{n}(g)\ket{\psi_{\rm g.i.}}=\ket{\psi_{\rm g.i.}}$, $\forall g \in G$ and $\forall n \in \Lambda$. 

\subsection{Ladder geometry and gauge fixing}

The formalism so far has been rather general. In the following, we specialize our discussion to a ladder geometry with open boundaries, which is the simplest nontrivial scenario for a pure gauge LGT.
We refer to the horizontal links of the ladder as leg links, denoted by $\ell_{\sigma}$, where $\sigma=\uparrow,\downarrow$ distinguishes the upper and lower leg. The vertical links connecting the two legs are called rungs and are denoted by $r$.
We implement a gauge fixing procedure to remove redundant degrees of freedom due to the local symmetries. 
In particular, we choose a gauge in which all variables on the legs of the ladder are set to the identity element $\ket{e}$. For the ladder geometry, this choice effectively reduces the system to a one-dimensional set of degrees of freedom residing on the rungs of the ladder. 
\begin{figure}[!htb]
    \centering
    \includegraphics[width=\linewidth]{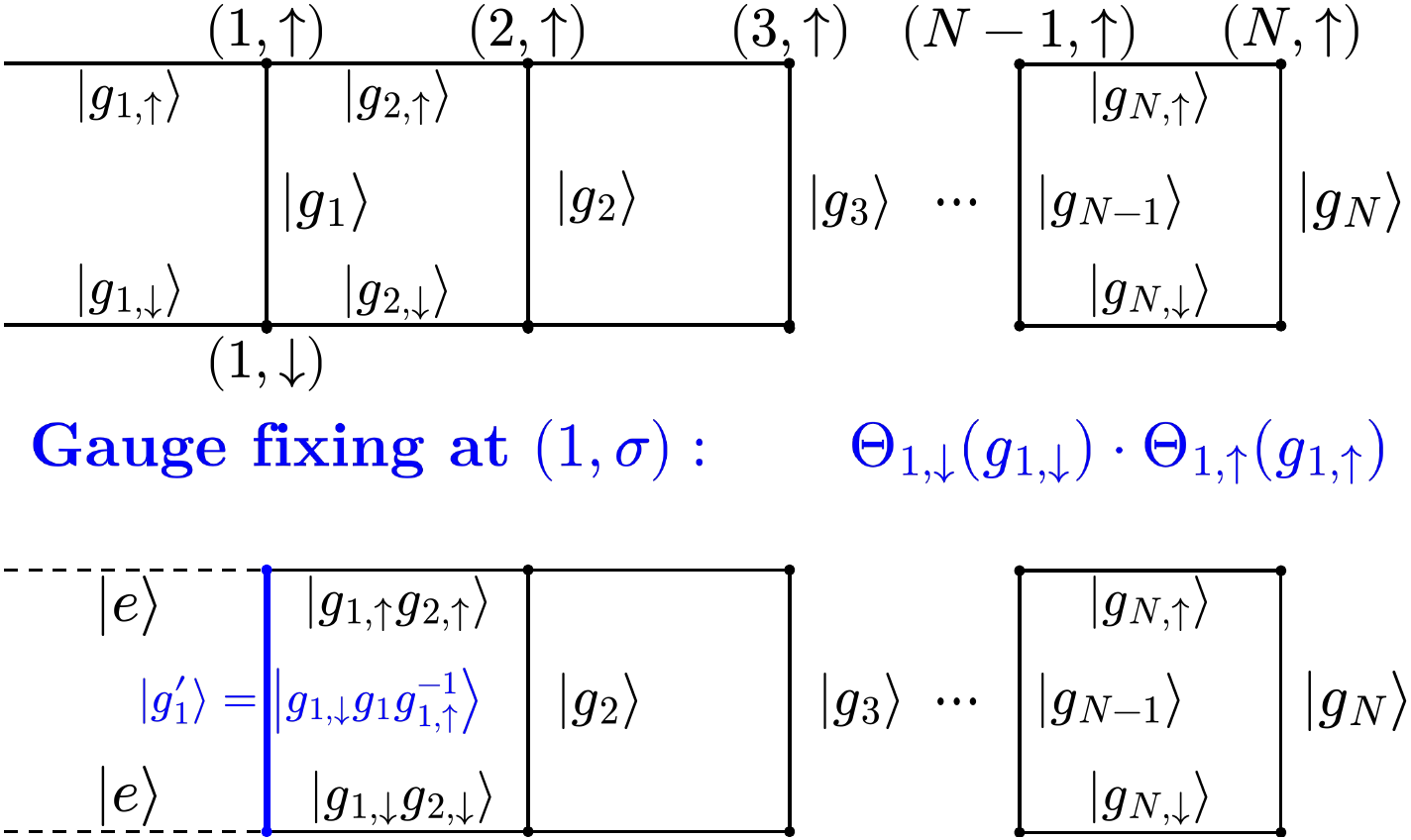}
    \caption{ \justifying Ladder with gauge degrees of freedom defined on all links (top) and the first step for the gauge-fixed version (bottom). At each vertex, a gauge transformation $\Theta_{n}(g)$ can be chosen that allows for setting a single leg variable to the identity, as illustrated for the vertices $(n,\sigma)$, $\sigma=\uparrow, \downarrow$. This process is iteratively applied through the entire ladder until both the upper and lower leg are in the identity group element. Only the rungs remain as dynamical degrees of freedom, at the cost of having introduced a string operator spanning the entire system that describes the non-Abelian electric-field energy of the legs. Further details are provided in Appendix~\ref{app:gauge_fixing}.}
    \label{fig:Gaugefixing}
\end{figure}

Let us consider a state $\ket{\vec{g}}$ in the group-element basis, $\ket{\vec{g}} \equiv \bigotimes_{\ell} \ket{g_\ell}$, where $g_\ell \in D_3$ labels the group element associated with link $\ell$ of the ladder. For a ladder of $N$ rungs, including the two dangling links at the left boundary, there are $3N$ links.
To eliminate the leg degrees of freedom, we apply local transformations $\Theta_{n}(g)$ at each vertex $n \in \Lambda$, with $g$ chosen to be the group element labeling the state of the leg on the left of the vertex $n$.  
Figure~\ref{fig:Gaugefixing} sketches this procedure for the first rung, which is presented in greater detail in Appendix~\ref{app:gauge_fixing}. 
An analogous scheme for SU(2) can be found, for instance, in Ref.~\cite{Dandrea2024}.
Besides removing the horizontal links, this gauge fixing also induces a corresponding transformation on the rung variables, which can be redefined thanks to the closure of the group. 
At the left boundary, we introduce two auxiliary dangling links, one for each vertex of the first rung. They ensure that Gauss's law is consistently imposed at the boundary without overconstraining the physical degrees of freedom on the first rung. The gauge transformation is therefore also applied at the two boundary vertices in order to fix the gauge degree of freedom on the dangling links.

The gauge fixing modifies the Kogut--Susskind Hamiltonian in Eq.~\eqref{eq:HKS}: the plaquette interaction becomes a two-body term and the electric field acting on the legs $H_-$ needs to be expressed in terms of rung variables only: 
\begin{equation}
    H = - \frac{1}{2\mathrm{g}^2} H_{\square} + \mathrm{g}^2 H_{|} + \mathrm{g}^2 (H^L_{-} + H^R_{-}) \ . 
\label{eq:Hfluxladder}
\end{equation}
Explicitly, the different terms are  
\begin{align}
    H_{\square} 
    &= \sum_{r=1}^{N-1} \mathrm{Tr}\!\left[U^{\dagger}_{r+1}U_{r}\right]+\mathrm{h.c.}\ ,
    \label{eq:Hmag} \\
    H_{|} 
    &= \sum_{r=1}^{N} \sum_{j} 
    \alpha_j \frac{\dim(j)}{|G|} 
    \sum_g \chi^j(g) \theta^L_r(g)\ ,
    \label{eq:Hele1} \\
    H^{L/R}_{-} 
    &= \scalebox{0.95}{$\displaystyle
    \sum_{r=1}^{N} \sum_{j} 
    \alpha_j \frac{\dim(j)}{|G|} 
    \sum_g \chi^j(g) \prod_{r'=1}^{r} \theta^{L/R}_{r'}(g) \ .
    $}
    \label{eq:Hele2}
\end{align}
Here, $r$ labels the rungs of the ladder and $\chi^j(g)=\mathrm{Tr}\,D^j(g)$ denotes the character of the group element $g$ in the irreducible representation $j$, of dimension $\dim(j)$. 

The Hamiltonian term in Eq.~\eqref{eq:Hmag} describes the gauge-fixed plaquette terms; since the horizontal legs of the ladder are set to the identity element of the group, plaquette operators are simplified to a two-body term acting on neighboring rungs. 
The second term $H_{|}$, Eq.~\eqref{eq:Hele1}, is the local electric term acting on the rungs, whose form remains unaffected by the gauge fixing.\footnote{The local projector on the $j$-th irrep is $\Pi_{r}^j=\frac{\dim (j)}{|G|}\sum_g \chi^j(g)\theta_r^L(g)$.} 
The last term $H^{L/R}_{-}$, Eq.~\eqref{eq:Hele2}, describes the electric field on the legs of the ladder; it explicitly contains string operators acting on all sites affected by the transformation required to fix the gauge on a specific leg. 
The detailed derivation of this term is presented in Appendix~\ref{app:leg_contribution}. 
The strength of both electric-field terms is determined by the coefficients $\alpha_j$, which depend only on the irreducible representation. For our computations with the $D_3$ gauge group, we fix $\alpha_j=-1,1,0$ for the identity ($\mathbf{1}$), parity ($\mathbb{P}$), and fundamental ($\mathbf{2}$) representations, respectively.
The corresponding irreducible representations and characters of \(D_3\) are summarized in Table~\ref{tab:D3_character_table}.
\begin{table}[h]
\centering
\begin{tabular}{c|ccc}
\hline\hline
Irrep & $\{e\}$ & $\{r,r^2\}$ & $\{s,sr,sr^2\}$ \\
\hline
$ \mathbf{1}$ & $1$ & $1$ & $1$ \\
$\mathbb{P}$ & $1$ & $1$ & $-1$ \\
$\mathbf{2}$ & $2$ & $-1$ & $0$ \\
\hline\hline
\end{tabular}
\caption{\justifying
Character table of $D_3$. The three conjugacy classes correspond to the identity, the two non-trivial rotations, and the three reflections.}
\label{tab:D3_character_table}
\end{table}

The gauge fixing procedure removes the local symmetries of the extended Hamiltonian; however, Eq.~\eqref{eq:Hfluxladder} still retains the invariance under a $D_3 \times D_3$ {\em global} transformation, generated by the string of left and right group multiplications
\begin{equation}
    \mathcal{Q}^L (g) = \prod_r \theta^L_r (g) \ , \quad \mathcal{Q}^R (g) = \prod_r \theta^R_r (g)\,.
    \label{eq:globalgaugetrans}
\end{equation}
This is a consequence of the two dangling legs at the boundary. Different symmetry sectors correspond to different $D_3$ static charges attached to these two legs. Equivalently, they identify distinct superselection sectors; as we will see below, the twisting of the magnetic interaction explicitly breaks this residual symmetry and thus couples these sectors.

\subsection{Tuning the degeneracies}
In the context of the ETH, degeneracies in the spectrum can deeply affect the model's behavior~\cite{Onoda2026, Santos2010, Xiaojun2023, Ebner2024, Rosenzweig1960, Giraud2022, Dabelow2026}.
The Hamiltonian in Eq.~\eqref{eq:Hfluxladder} exhibits a highly degenerate spectrum. Part of these degeneracies arises from symmetries of the Hamiltonian: the vertical inversion of the rung variables, a spatial symmetry, and the residual $D_3 \times D_3$ global gauge transformations in Eq.~\eqref{eq:globalgaugetrans}, which are internal symmetries. These degeneracies can be resolved by working in fixed symmetry sectors, at the price of considering only a subspace of the total Hilbert space. 

Inspired by Ref.~\cite{Munk2018}, we take an alternative route and add a controllable deformation of the magnetic interaction, in the gauge-fixed frame, that mixes the superselection sectors. This allows us to consider the full spectrum of the Hamiltonian and resolve only the vertical inversion symmetry $\mathcal{I}U_r \mathcal{I}^\dagger =U_r^\dagger$. Another consequence is that the ground state of the deformed Hamiltonian at $\mathrm{g}\to 0$ becomes a product state where all links are in the group identity, instead of a $\mathbb{Z}_6$-symmetric GHZ state.
Studying how the effective thermalization properties, in particular the emergent GOE level statistics, depend on the deformation strength and the system size allows us to estimate the behavior the clean LGT model would take at sizes beyond our numerical reach.

The deformation is implemented through a rung-dependent unitary transformation of the operator $U^j$, 
\begin{equation}
    U^j_r\rightarrow T_r U^j_r T^{\dagger}_r \ .
\end{equation}
Under this transformation, the plaquette term in Eq.~\eqref{eq:Hmag} becomes
\begin{equation}
    \mathrm{Tr}\left[T^{\dagger}_{r+1} T_r U_r T^{\dagger}_r T_{r+1} U^{\dagger}_{r+1}\right] \ .
    \label{eq:Hmagmod}
\end{equation}
We choose the unitary matrices $T_r$ such that the relative transformation between neighboring rungs is independent of $r$, i.e., $T_{r+1}^\dagger T_r=C$. 

Here, following Ref.~\cite{Munk2018} we use \footnote{This form of $C$ is only strictly unitary under a normalization with $\sqrt{(1+\varepsilon^2)/2}$, which corresponds to simply rescaling the Hamiltonian parameters. 
We neglect this subtlety here, as it does not affect the thermalization properties of the model. 
Unitarity of $C$ in the present form is recovered at $\varepsilon=\pm 1$ and in the limit of small $\varepsilon$, where it becomes equivalent to setting $T_r = e^{ir\phi}e^{-ir\varepsilon D}$,  with $D = \frac{1}{\sqrt{3}} (\sigma_x +\sigma_y -\sigma_z) $. }
\begin{equation}
    C = \frac{e^{-i\pi/4}}{\sqrt{2}} \bigg[ \mathbb{I} +\frac{i \varepsilon}{\sqrt{3}} (\sigma_x +\sigma_y -\sigma_z) \bigg] \ .
\end{equation} 
The parameter $\varepsilon$ plays the role of a chiral deformation parameter. It controls the departure of $C$ from a matrix proportional to the identity, and therefore sets the strength with which the modified plaquette term breaks the residual global gauge symmetry.

Applying the global left and right gauge transformation operators, defined in Eq.~\eqref{eq:globalgaugetrans}, to the plaquette operator, we get
\begin{equation}
\begin{split}
    & \mathcal{Q}^{L\dagger}(g) \mathrm{Tr}\left[C^{\dagger} U^{\dagger}_{r+1}C U_{r}\right] \mathcal{Q}^L(g)  \\
    & = \mathrm{Tr}\left[C^{\dagger}U^{\dagger}_{r+1}D^{\dagger}(g)CD(g) U_{r}\right] \\
    & \neq \mathrm{Tr}\left[C^{\dagger}U^{\dagger}_{r+1}C U_{r}\right] \ ,
\end{split} 
\end{equation}
\begin{equation}
\begin{split}
    &\mathcal{Q}^{R \dagger}(g) \mathrm{Tr}\left[C^{\dagger} U^{\dagger}_{r+1}C U_{r}\right] \mathcal{Q}^R (g) \\
    & = \mathrm{Tr}\left[C^{\dagger}D^{\dagger}(g)U^{\dagger}_{r+1}C U_{r}D(g)\right] \\
    & \neq \mathrm{Tr}\left[C^{\dagger} U^{\dagger}_{r+1}C U_{r}\right] \ .
\end{split} 
\end{equation}
From these equations, we see that the new plaquette term in Eq.~\eqref{eq:Hmagmod} is not invariant under the global gauge transformations, unless $C$ commutes with all matrices $D(g)$ of the irreducible representation. By Schur's lemma, this is possible only for $C \propto \mathbb{I}$. Therefore, any nontrivial choice of $C$ explicitly breaks the residual global gauge symmetry.
This symmetry breaking provides the analogue, in the present non-Abelian setting, of the chiral interaction appearing in the chiral Potts model~\cite{Howes1983,Fendley_2012,Moran_PRB2017,Munk2018,Wauters2025}. 
In an Abelian group, such as $\mathbb{Z}_N$, $C$ reduces to a phase that is responsible for the emergence of helical phases.
In $D_3$, which lacks the cyclic structure of $\mathbb{Z}_N$, the matrix $C$  deforms the plaquette operator breaking its invariance under conjugation. Instead of the standard magnetic ordering, i.e., vanishing flux, its energy is minimized by having both rungs in the identity element. 

\section{Thermalization properties}\label{sec:ETH}
In this section, we numerically investigate the thermalization properties of the Hamiltonian in Eq.~\eqref{eq:Hfluxladder}.
All results are obtained through exact diagonalization (ED) of ladders with up to $N=5$ rungs.\footnote{In terms of Hilbert space dimension, this corresponds to $\dim \mathcal{H}=7776$, which is comparable to full ED on a 13-qubit Hamiltonian. Increasing the chain length to 6 rungs leads to $\dim \mathcal{H}=46656$, comparable to 15-16 qubits. Without gauge fixing, we would have $\dim \mathcal{H}\sim 1.3\times 10^{10}$ for 5 rungs.}
Here, we focus on two complementary probes of ETH: the energy gap statistics, as well as the diagonal and off-diagonal matrix elements of local observables in the Hamiltonian basis.

\subsection{Spectral statistics}
We start by analyzing the model's spectral properties, which are among the most established signatures of the emergence of quantum chaos and thermalization in closed quantum systems~\cite{Berry1977, Bohigas1984, Mehta2004, Oganesyan2007, Atas2013, D_Alessio2016}.
Before computing level-spacing statistics, we resolve the exact vertical-inversion symmetry of the gauge-fixed ladder $\mathcal{I}$. Unless otherwise stated, all gap-ratio results are obtained in the even sector $\mathcal{I}=+1$.
Through the comparison of the numerical results with the Poisson and Gaussian orthogonal ensemble (GOE) predictions, we aim to identify regions of parameter space where they suggest thermalization is suppressed and where chaotic behavior emerges.

A figure of merit to discriminate between these different behaviors is the minimal gap ratio $r=\{r_n\}_{n=1}^{D-2}$~\cite{Oganesyan2007}, where $D = \dim \mathcal{H}$ is the dimension of the Hilbert space under consideration. From each energy gap pair between consecutive levels ($E_{n+1} -E_n,\ E_n - E_{n-1}$), it is defined as
\begin{equation}
    r_n = \min\left(\frac{E_n-E_{n-1}}{E_{n+1}-E_{n}},\frac{E_{n+1}-E_{n}}{E_n-E_{n-1}}\right) \ .
\end{equation}
In contrast to the bare level-spacing statistics, the minimal gap ratio does not require an unfolding procedure to flatten the density of states~\cite{Oganesyan2007}.
According to the random matrix theory of Hermitian matrices, this quantity distinguishes chaotic from integrable behavior: in the first case, the system exhibits level repulsion and the distribution of $r$ is peaked away from zero, approaching one of the Gaussian ensembles depending on the Hamiltonian symmetries; in the second, the energy levels are uncorrelated and, in the presence of sufficient disorder, the minimal gap ratio follows a Poisson distribution peaked at zero~\cite{Berry1977,Oganesyan2007,Atas2013}. 
Finite-size effects and unresolved symmetries typically lead to deviations from these known limits.

To characterize the spectral statistics at different values of the coupling $\mathrm{g}$ and of the chiral deformation parameter $\varepsilon$, we compute the average minimal gap ratio $\braket{r} = \frac{1}{D-2}\sum_{n=1}^{D-2} r_n$.
Figure~\ref{fig:spectralstat}(a) shows $\braket{r}$ in the parameter space for $N=5$.  
The shaded areas indicate the GOE-compatible region for different system sizes, identified by $\braket{r}\simeq \braket{r}_{\rm GOE}=0.53$~\cite{Atas2013}. This regime, already visible for $N = 3$, becomes progressively more extended as the system size increases up to $N = 5$. This finite-size trend suggests an increasingly robust chaotic regime.
\begin{figure}
    \centering
    \includegraphics[width=\linewidth]{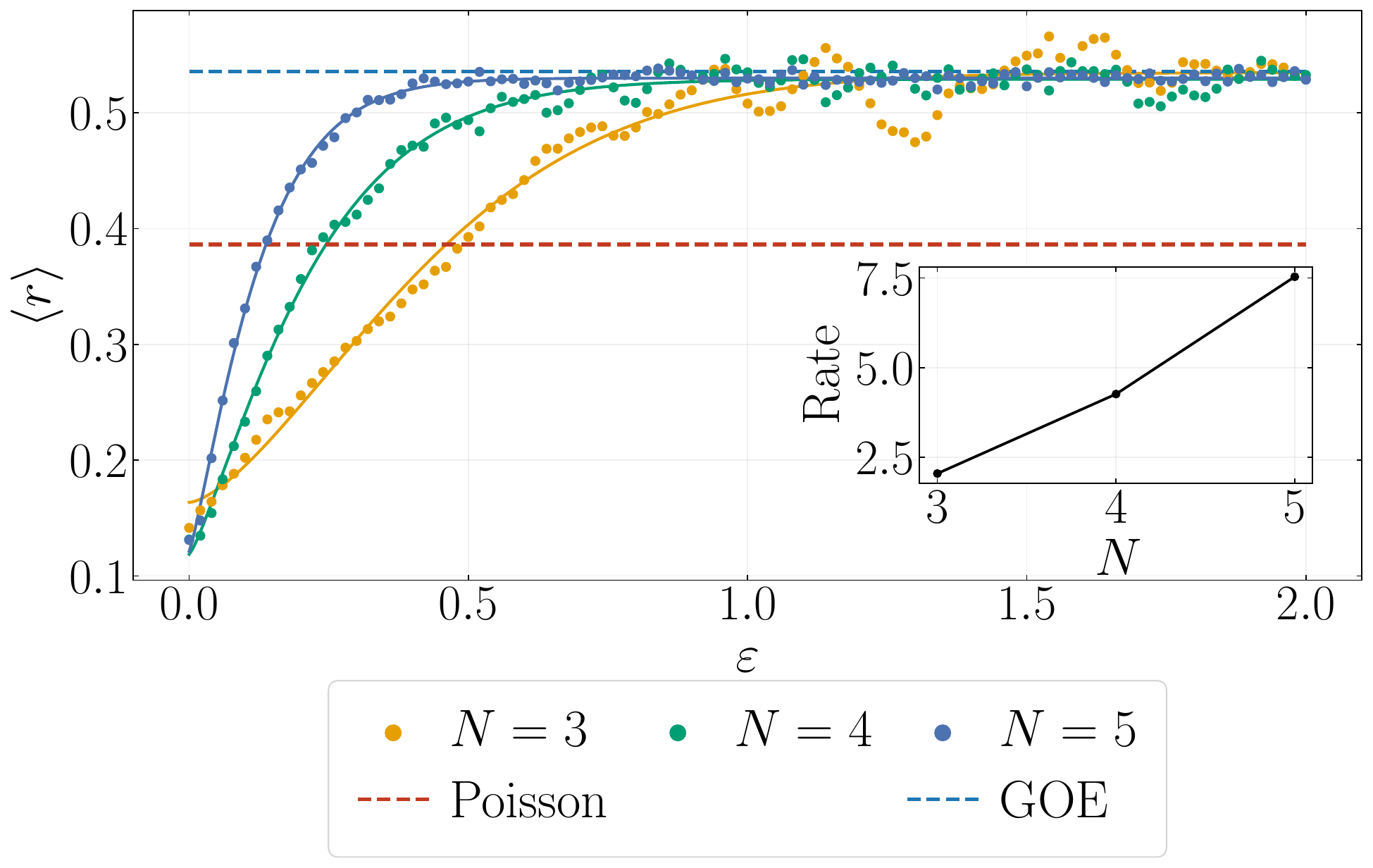}
    \caption{\justifying
    Average gap ratio as a function of the deformation parameter $\varepsilon$, for $N=3,4,5$ and fixed coupling $\mathrm{g}=1.02$. The solid lines are stretched-exponential fits $\braket{r}=\braket{r}_{\rm GOE}(1-Ae^{-b \varepsilon^c})$.  The horizontal dashed lines indicate the Poisson and GOE reference values. As the system size increases, the crossover toward GOE statistics occurs at smaller values of $\varepsilon$, showing that the chaotic regime becomes increasingly robust. The inset reports the fitted rate parameter as a function of $N$.}
    \label{fig:gap_ratio_fixed_g}
\end{figure}

At the considered system size, a region with a level-spacing distribution that significantly deviates from GOE statistics is also present. A point in this regime is shown in Fig.~\ref{fig:spectralstat}(b), where the gap-ratio distribution lies rather close to the Poisson one $P_{\rm Poisson}(r) = \frac{2}{(1+r)^2}$~\cite{Atas2013}, suggesting the absence of level repulsion. 
However, this qualitative agreement is an accidental feature of the crossover from the degenerate spectrum to the chaotic one. In the thermodynamic limit, one can expect clear chaotic signatures in the spectrum for arbitrarily small non-integrable perturbation~\cite{Santos2010}.
This statement is supported by the finite-size trend of Fig.~\ref{fig:spectralstat}(a) and Fig.~\ref{fig:gap_ratio_fixed_g}, where the GOE-compatible region expands with $N$ and the residual non-chaotic window shrinks toward the undeformed point $\varepsilon = 0$.  
Furthermore, even the deformation-free model itself should not be expected on physical grounds to be integrable, since its plaquette term is reminiscent of a non-Abelian generalization of the Ising/Potts model in a longitudinal field, suggesting that the entire parameter regime of finite $\mathrm{g}$ and $\varepsilon$ will show thermalizing behavior in the thermodynamic limit.

In contrast, inside the chaotic region, the match between the numerical data and the theoretical prediction is robust and works well everywhere in the shaded area. Figure~\ref{fig:spectralstat}(c) shows data for an example point, where the distribution follows the Wigner--Dyson GOE prediction $P_{\rm GOE}(r) = \frac{27}{4}\frac{r(r+1)}{(1+r+r^2)^\frac{5}{2}}$~\cite{Atas2013}.

By choosing a fixed coupling, we can follow the crossover more directly as a function of the deformation parameter $\varepsilon$, as shown in Fig.~\ref{fig:gap_ratio_fixed_g}. 
For all system sizes, as $\varepsilon$ is increased, the average minimal gap ratio converges exponentially fast toward the GOE value. 
The crossover becomes sharper and is shifted toward smaller values of $\varepsilon$ as $N$ increases.
The fitted rate parameter provides a useful qualitative measure of this sharpening; over the small sizes we consider, it displays a growth that is at least linear in $N$, see the inset of Fig.~\ref{fig:gap_ratio_fixed_g}.
In summary, this behavior supports the interpretation that the non-chaotic region progressively shrinks to disappear for sufficiently large $N$, while the chaotic regime becomes more robust with increasing system size.

\subsection{Local observables}
The spectral properties suggest that the $D_3$ LGT on a ladder exhibits chaotic behavior in extended parameter regimes. We now test numerically whether local operators $\hat{O}$ within that region follow the ETH. 
In case they do, their matrix representation in the Lehmann basis $\{|n\rangle\}$ can be written as~\cite{D_Alessio2016} 
\begin{equation}
    O_{mn} = O(\overline{E}) \delta_{mn}+e^{-S(\overline{E})/2}f_O (\overline{E}, \omega) R_{mn} \ ,
    \label{eq:ETH_general}
\end{equation}
where $O(\overline{E})$ is a smooth function of the mean energy $\overline{E} \equiv (E_m + E_n)/2$ and corresponds to the microcanonical expectation value. 
The second term describes eigenstate-to-eigenstate fluctuations around this smooth behavior. The function $f_O(\overline{E},\omega)$ is the spectral response function of the operator $\hat{O}$. It is assumed to be smooth in both the average energy $\overline{E}$ and the energy difference $\omega \equiv E_n-E_m$. 
The overall magnitude of the fluctuations is suppressed by the entropic factor $e^{-S(\overline{E})/2}$, where $S(\overline{E})$ is the microcanonical entropy associated with the density of states. The fluctuations are encoded in the random variable $R_{mn}$, which in the ETH regime is expected to be normally distributed with zero mean and unit variance.
\begin{figure*}[!htb]
    \centering
    \includegraphics[width=\linewidth]{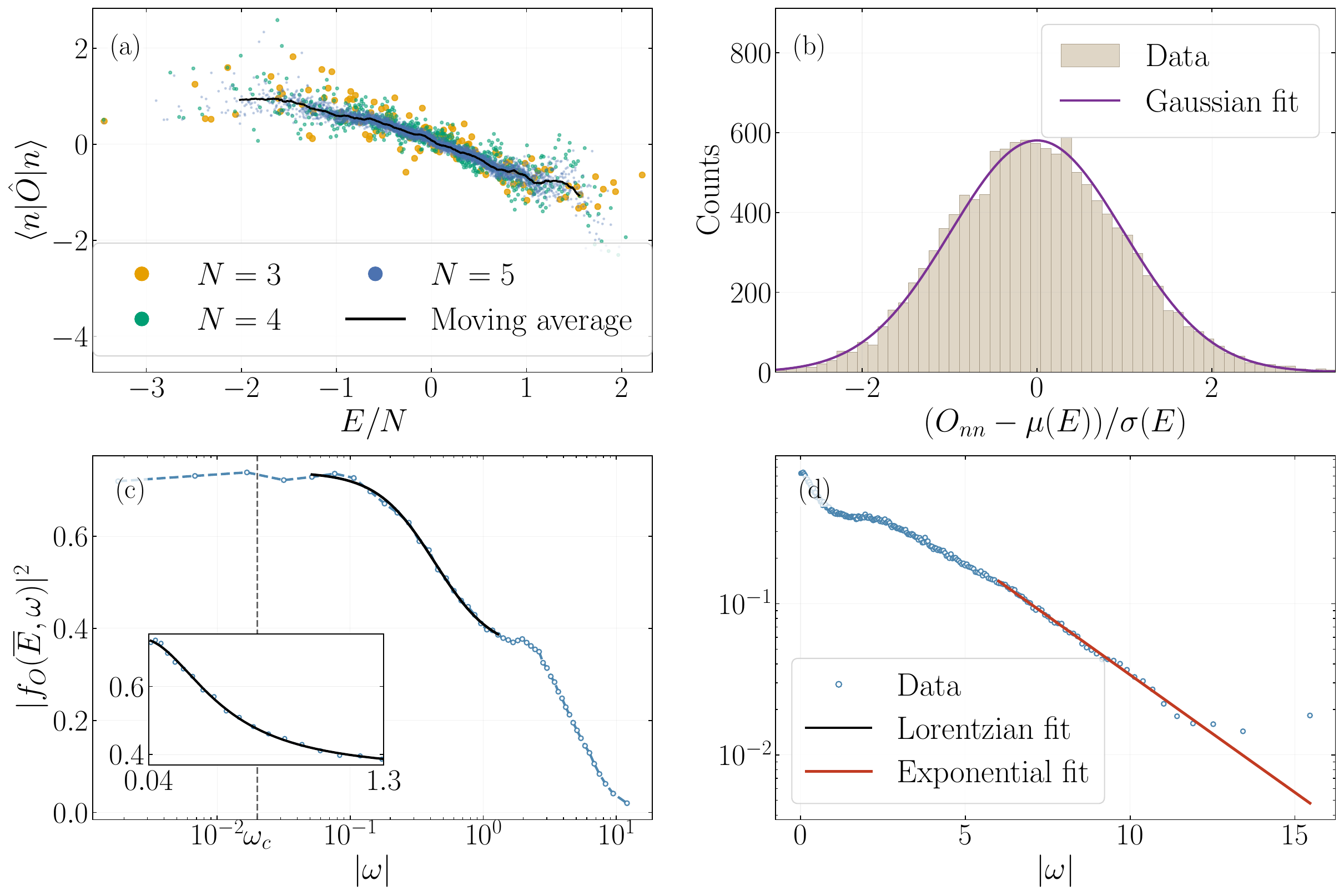}
    \caption{ \justifying 
    Effective thermalization of the plaquette operator $\hat{O}$ at a representative point with GOE level statistics with $\mathrm{g}=1.06$ and $\varepsilon=1.24$. 
    (a) Diagonal matrix elements $O_{nn}=\langle n|\hat{O}|n\rangle$ of the single-plaquette operator as a function of the energy density $E/N$. 
    The data spread decreases for increasing system size, consistent with ETH, approaching the smooth running average, shown here for $N=5$  (black line). 
    (b) Normalized residuals $(O_{nn}-\mu(E))/\sigma(E)$ obtained from the $N=5$ data in panel (a). The data are well approximated by a Gaussian fit (purple solid line).
    (c-d) Frequency dependence of the function $|f_O(\overline{E},\omega)|^2$ extracted from the off-diagonal matrix elements $\bra{n}\hat{O}\ket{m}$, at fixed mean energy $\overline{E}$, for $N=5$.
    In (c), the log-scale on the frequency axis emphasizes the expected plateau at small $|\omega|$ and the Lorentzian decay (inset) over the range $|\omega|\in[0.04,1.3]$. 
    In (d), we highlight the exponential tail over the range $|\omega|\in[6.0,16.0]$. In panel (c), adaptive frequency bins are used to enhance the resolution at low frequencies, whereas panel (d) employs approximately equal-population bins.
   }
    \label{fig:diagonal_eth}
\end{figure*}

We start by characterizing the diagonal matrix elements of the plaquette operator 
\begin{equation}
    \hat{O} = 2\,\Re \left(\mathrm{Tr}\!\left[ C^\dagger U_{r+1}^{\dagger} C U_r\right] \right) \ .
\end{equation}
This Hermitian operator coincides with the local contribution entering the deformed magnetic Hamiltonian, up to the overall coupling prefactor. To reduce boundary effects, we choose the plaquette closest to the center of the ladder, with index $r=\lfloor N/2\rfloor$.

The numerical results for the point in parameter space exemplified in Fig.~\ref{fig:spectralstat}(c), where we have identified a chaotic level-spacing statistics, are displayed in Fig.~\ref{fig:diagonal_eth}(a); it reports the diagonal ensemble $\bra{n}\hat{O}\ket{n}$ as a function of the energy density $E/N$ for system sizes $N=3,4,5$. 
The data fully confirms the behavior expected from the ETH ansatz in Eq.~\eqref{eq:ETH_general}. 
The diagonal matrix elements become increasingly smooth functions of energy as the system size grows, except near the spectral edges, and the spread between eigenstates with nearby energies is reduced with increasing system size.
In Fig.~\ref{fig:diagonal_eth}(b), we show the normalized fluctuations of the diagonal matrix elements around their local mean. The data are obtained by sampling random energy windows of width $\Delta E=0.2$ within the bulk of the spectrum. The residual fluctuations are well described by Gaussian statistics, in accordance with the ETH expectation.

The off-diagonal ETH ansatz is characterized by the function $f_O(\overline{E}, \omega)$, which is closely related to the spectral function of the operator $\hat O$ and to the dissipative part of the corresponding linear-response susceptibility. 
It encodes information on various time scales associated with the operator thermalization~\cite{D_Alessio2016}.
From Eq.~\eqref{eq:ETH_general}, one obtains that $|f_O(\overline{E},\omega)|^2$ is proportional to the average squared off-diagonal matrix elements of the operator $\hat O$. 
More explicitly, for each pair of eigenstates we compute the mean energy $\overline{E}_{mn}$, the energy difference $\omega_{mn}$ and the off-diagonal expectation value $O_{mn}$. We then restrict the data to an energy window
$|\overline{E}_{mn}-\overline{E}_0|<0.4$, where $\overline{E}_0$ is chosen at the maximum of the density of states in the even inversion sector $\mathcal{I}=+1$. 
Within this energy window, the selected off-diagonal matrix elements
are binned according to their frequency $\omega_{mn}$. For each frequency bin, we estimate the squared spectral function as $|f_O(\overline{E}_0,\omega)|^2 \simeq e^{S(\overline{E}_0)} \langle |O_{mn}|^2\rangle_{\omega}$, where the average is performed over all pairs $(m,n)$ inside the same $\omega$-bin and satisfying the constraint on the average energy. 
The entropy $S(E)$ is estimated from the numerical density of states as $S(E) = \log \rho(E)$. Therefore, data plotted in Fig.~\ref{fig:diagonal_eth}(c-d) correspond to the entropy-rescaled mean-square amplitude of the off-diagonal matrix elements at fixed average energy.

The numerical results are shown in Fig.~\ref{fig:diagonal_eth}(c-d). The low-frequency behavior is well fit by a phenomenological Lorentzian form~\cite{Ebner2024,Schonle2021}\footnote{For the available data ranges, other fit functions also yield reasonable results.}
\begin{equation}\label{eq:lorentzian}
    |f(|\omega|)|^2 = \frac{a}{\omega^2+b}+c \ . 
\end{equation}
Since $\overline{E}_0$ lies at the maximum of the density of states, $f(|\omega|)$ is effectively probing the susceptibility at infinite effective temperature. Hence, for sufficiently small $\omega$, $|f_O(\overline{E}_0,\omega)|^2$ is directly the Fourier transform of the connected autocorrelation function of $\hat O$~\cite{D_Alessio2016,Srednicki1999}. The Lorentzian component of Eq.~\eqref{eq:lorentzian} therefore corresponds to an exponential decay of this autocorrelation, with relaxation rate $\sqrt{b}$~\cite{Schonle2021}. 

At intermediate frequencies, nonuniversal features typically appear, whose shape and positions reflect the properties of specific model excitations.
For pure gauge non-Abelian theories, for instance, quasi-particle excitations are typically related to glueballs, although their persistence at high energies depends, i.a., on the strength of the coupling~\cite{Ebner2024}.
In the present case, the combination of the magnetic deformation and relatively weak coupling, close to the electric-magnetic crossover, prevents the emergence of clear quasiparticle excitations: in our data, the intermediate-frequency features appear only as a ``knee'' in $|f_O(\overline{E}, \omega)|^2$, which becomes a broad feature if the local operator $\hat{O}$ is a plaquette on the boundary of the ladder (see Appendix~\ref{app:boundary_plaquette}).

Finally, at large $\omega$, we note a fast decay of $|f_O(\overline{E}, \omega)|^2$. In general, for lattice Hamiltonians with local, bounded terms, the matrix elements of local operators between energy eigenstates are bounded by a function decaying exponentially in $|\omega|$~\cite{D_Alessio2016,Arad_2016}. This applies directly to the undeformed model, which is local in the original Kogut--Susskind formulation and gauge fixing is a unitary change of basis that leaves the matrix elements of gauge-invariant operators unchanged, see Appendix~\ref{app:gauge_fixing}. 
For $\varepsilon\neq0$, the deformation is defined in the gauge-fixed Hamiltonian, which contains the non-local string operators of Eq.~\eqref{eq:Hele2}, so the bound does not immediately apply. For the available data, the high-frequency tail of Fig.~\ref{fig:diagonal_eth}(d) nevertheless shows a rapid decay, which is well fit by an exponential. The accessible system sizes do not allow us to exclude logarithmic corrections, which would be in agreement with the generic expectation for low-dimensional chaotic many-body systems~\cite{Parker_PRX2019,Avdoshkin2020}.

\section{Entanglement structure}
\label{sec:entanglement}
In this section, we focus on non-local spectral properties and analyze ETH through the lens of the entanglement entropy and the dynamics of the entanglement Hamiltonian.
The first provides information about the structure of the eigenstates, while the second quantifies the effective thermalization of the reduced density matrix. 

\subsection{Entanglement entropy}
Thermalization properties of quantum systems are related to their entanglement structure, which can be probed, for instance, by the bipartite entanglement entropy. 
In non-integrable models, eigenstates close to the spectral edges are expected to retain relatively low entanglement, while excited states in the bulk of the spectrum typically display volume-law entanglement~\cite{Miao2021, Vidmar2017, Geraedts2016}. 
Within the eigenstate thermalization hypothesis, the entanglement entropy is expected to vary smoothly with energy and to be related, up to finite-size corrections, to the von Neumann entropy of the corresponding thermal ensemble~\cite{Deutsch1991,D_Alessio2016}.
However, violations of this general picture may occur in the form of a few isolated excited states with anomalously low entropy; these are known as quantum scars~\cite{Bernien2017,Serbyn2021,Moudgalya_2022} and have attracted considerable interest in recent years in the context of LGTs~\cite{Surace_PRX2020,Banerjee2021,Saptarshi2022,Aramthottil2022,Halimeh_2023, Hayata2023_1,Sau2024,Ge2024,Calajo2025}.

\begin{figure}[!htb]
    \centering
    \includegraphics[width=\linewidth]{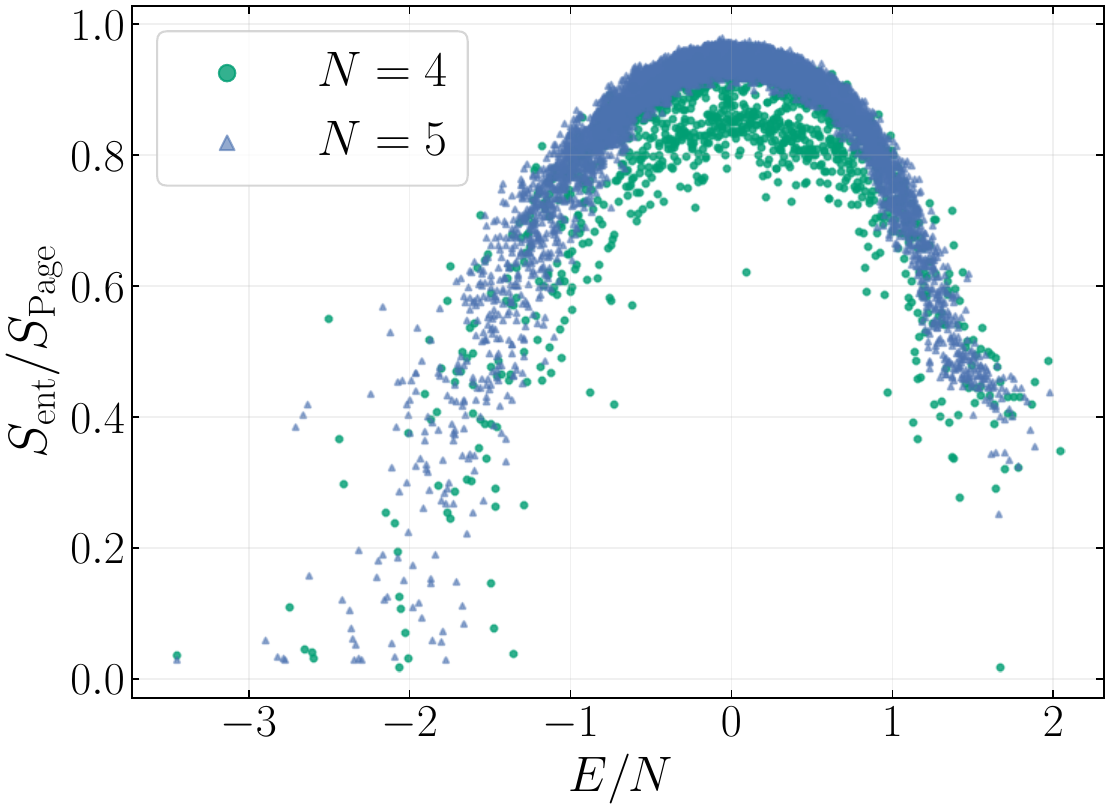}
    \caption{ \justifying Entanglement entropy $S_{\rm ent}$ as a function of the energy density $E/N$ for parameters $\mathrm{g}=1.06$ and $\varepsilon=1.24$, shown for system sizes $N=4$ and $N=5$. Eigenstates form a smooth dome-shaped structure, consistent with the expected thermal behavior of the bulk of the spectrum.}
    \label{fig:entropy}
\end{figure}

To understand whether the discrete non-Abelian gauge theory given by Eq.~\eqref{eq:Hfluxladder} follows the standard pattern for non-integrable models or hosts scar states, we analyze the bipartite entanglement entropy $S_{\rm ent}\equiv S_{\rm ent}(\left\lfloor N/2\right\rfloor)$ of the Hamiltonian eigenstates as a function of the energy. 
Figure~\ref{fig:entropy} shows $S_{\rm ent}$, normalized by the Page value $S_{\mathrm{Page}}$ corresponding to the chosen bipartition~\cite{Page1993}, as a function of the energy density $E/N$, in the chaotic regime.
Consistent with the ETH expectation, for both $N=4,5$ the majority of the eigenstates form a smooth dome-shaped structure, with maximal entanglement in the middle of the spectrum and smaller values near the spectral edges.

Our results are in line with similar analysis on SU(2) pure gauge theories, where no violations of thermalization have been reported~\cite{Xiaojun2023,Ebner2025}. 
Instead, it contrasts with truncated pure gauge U(1) LGTs, where scar states emerge~\cite{ Banerjee2021, Saptarshi2022, Sau2024}. 
In non-Abelian LGTs, the freedom granted by the internal structure of bigger than one-dimensional irreps may boost the thermal behavior and hinder the occurrence of scar states.
\begin{figure*}[!htb]
    \centering
    \includegraphics[width=\textwidth]{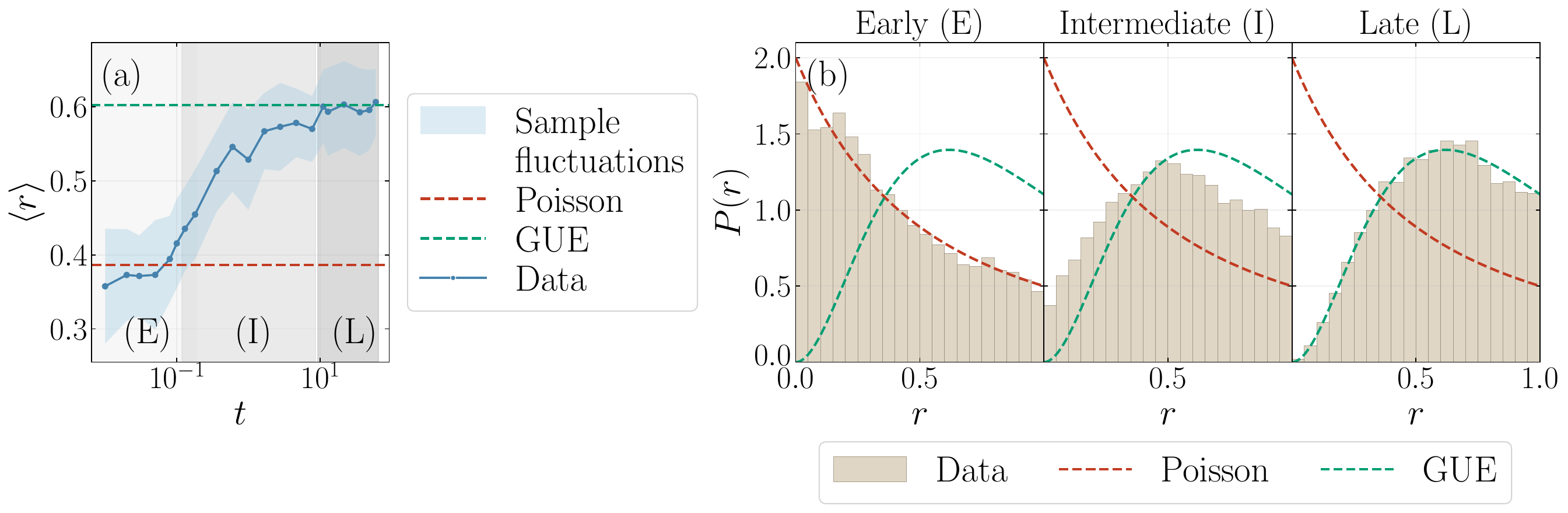}
    \caption{ \justifying
    (a) Time evolution of the average gap ratio of the entanglement spectrum, averaged over $100$ initial eigenstates. These are obtained from $100$ different initial Hamiltonians and randomly selected eigenstates that satisfy the imposed initial-state criteria. All states are evolved with the same final quench Hamiltonian with $\mathrm{g}=1.3$ and $\varepsilon=1.0$. The gap-ratio statistics are computed within a fixed $\mathcal I_A$ sector and in the bulk of the entanglement spectrum, discarding the lowest and highest $10 \%$ of the entanglement energies to reduce edge effects. The horizontal lines indicate the reference values for Poisson statistics (red dashed) and GUE statistics (green dashed). The shaded region indicates the standard deviation across the sample. 
    One can discern Early (E), Intermediate (I), and Late (L) regimes, where the EH level statistics, respectively, lie close to Poissonian, cross over, and approach the GUE value. 
    (b) Distribution of the gap ratios, collected over the initial states and over all times belonging to the E, I, and L regimes. In the E regime, the distribution follows a Poisson curve (red dashed) rather closely, and in the L regime it is well approximated by a GUE distribution (green dashed).}
    \label{fig:entspectrum}
\end{figure*}
\subsection{Entanglement Hamiltonian}
Entanglement is also useful to monitor the thermalization through the system dynamics.
In particular, we focus on the spectral statistics of the entanglement Hamiltonian (EH)~\cite{Yang_2017, Li2008,Garrison2018,Dalmonte2022}, which has been shown to be a robust probe of quantum chaos~\cite{Geraedts2016,Mueller2022,Mueller2025}.
Indeed, the statistical properties of the entanglement spectrum provide information on how (local) thermalization emerges from the structure of the reduced density matrix.
Tracking the time evolution of the entanglement spectrum during a quench dynamics provides a powerful framework for characterizing thermalization, complementary to standard ETH analysis.

We consider the time evolution of an initial state $\rho_0=\ket{\Psi_0}\bra{\Psi_0}$ after a quantum quench and study a subsystem $A$ of $N_A \leq N/2$ rungs. Its reduced density matrix $\rho_A(t)$ after time $t$ is given by
\begin{equation}
    \rho_A(t) \equiv \mathrm{Tr}_{\bar{A}}[\rho(t)] \ , \quad \rho(t)\equiv \mathcal{U}(t) \rho_0 \mathcal{U}(t)^\dagger \ ,
\end{equation}
where $\bar{A}$ is the complement of $A$, and $\mathcal{U}(t)=\mathrm{exp}(-i H t)$ is the time evolution operator generated by the post-quench Hamiltonian $H$. 
The associated entanglement Hamiltonian is defined as 
\begin{equation}\label{eq:EH_def}
    H_{\mathrm{ent}} (t) \equiv - \log(\rho_A(t)) \ .
\end{equation}
Thermalization properties are revealed by the statistics of the EH's spectrum $ \{ \xi_{\lambda} \}$, called the entanglement spectrum, which can be retrieved from the spectral decomposition of the reduced density matrix $\rho_A(t) \equiv \sum_{\lambda} p_{\lambda} \ket{\lambda(t)} \bra{\lambda(t)}$. The EH's eigenvalues are then defined as {$ \{ \xi_{\lambda} (t) \equiv - \log(p_{\lambda}(t)) \}$}. 

Since the Hamiltonian preserves the inversion symmetry \(\mathcal I\), the reduced density matrix \(\rho_A(t)\) inherits a corresponding subsystem symmetry \(\mathcal I_A\). 
As a consequence, the entanglement Hamiltonian is block diagonal in the eigenspaces of \(\mathcal I_A\). 
We therefore compute the entanglement-spectrum gap ratio after resolving these symmetry sectors. 
In practice, we diagonalize \(\rho_A(t)\) separately in the \(\mathcal I_A=\pm1\) sectors and compute the minimal gap ratios within each block. 
This avoids artificially mixing independent symmetry sectors, which would suppress the average gap ratio.

To implement this diagnostic, we monitor the post-quench time evolution of the entanglement spectrum for a system of $N=5$ rungs, which we bipartition into a subsystem \(A\) containing the first $N_A=2$ rungs and its complement. We sample an ensemble of $100$ initial states by considering a $10\times 10$ grid of initial Hamiltonians with parameters $(\mathrm{g}_i,\varepsilon_i)$, where $\mathrm{g}_i\in[0.25,0.35]$ and $\varepsilon_i\in[0.15,0.25]$. 
These belong to the parameter region in Fig.~\ref{fig:spectralstat} where for $N=5$ the static level-spacing statistics deviates significantly from Wigner--Dyson.
For each initial Hamiltonian, we diagonalize the spectrum and identify a set of candidate eigenstates with comparable post-quench energy and controlled initial entanglement properties. 
We then randomly select one eigenstate from this set. All selected states are evolved with the same final Hamiltonian, with parameters $\mathrm{g}=1.3$ and $\varepsilon=1.0$. This protocol avoids relying on a single fine-tuned initial condition, while keeping the ensemble sufficiently homogeneous to compare the resulting entanglement-spectrum dynamics.

In Fig.~\ref{fig:entspectrum}(a), we show the average gap ratio $\langle r\rangle$ of the EH as a function of the evolution time $t$, averaged over the $100$ randomly chosen initial states. 
The time evolution shows a clear buildup of level repulsion. The averaged gap ratio $\langle r\rangle$ evolves from $\approx 0.4$ toward the Gaussian-unitary-ensemble (GUE) value $\braket{r}_{\rm GUE}\approx 0.6$ , characteristic of repulsive level statistics~\cite{Atas2013}.  
As a side remark, the random matrix ensembles describing the entanglement spectrum and the Hamiltonian spectrum are different. The Hamiltonian of our system has time-reversal symmetry ($H$ is real), hence leading to a GOE distribution, while the reduced density matrix is typically complex, thus having a GUE level statistics. 

From the numerical data, we identify three regimes: an early-time regime (E), where the statistics remain close to the initial distribution; an intermediate regime (I), where $\langle r\rangle$ gradually crosses over to the GUE value; and a late-time regime (L), where it saturates around the GUE prediction.
To confirm that this behavior is reflected in the full statistics of the minimal gap ratio,
Fig.~\ref{fig:entspectrum}(b) reports the normalized gap-ratio distributions $P(r)$ for the three ranges of evolution times.
The statistics evolve from an early-time (E) Poisson-like distribution, with no level repulsion, to a late-time (L) distribution compatible with the GUE prediction. The intermediate-time (I) distribution lies between these two limits.  
Within the sampled initial eigenstates, the level statistics of the entanglement spectrum are found to be largely independent of the choice of the initial state. 

Overall, the late-time behavior indicates that thermalization is reflected not only in local observables but also in the spectral correlations of the reduced density matrix. 
Furthermore, it provides a more robust probe of dynamical thermalization than local observables for small systems, as the latter are more sensitive to finite-size effects.

\section{Conclusion}\label{sec:conclusions}
In this work, we investigated thermalization in an isolated quantum many-body system with a discrete non-Abelian local gauge symmetry $D_3$. 
Since gauge constraints strongly restrict the accessible Hilbert space, it is not obvious whether standard signatures of quantum chaos and thermalization should emerge as in unconstrained systems. 
To address this question, we studied a pure gauge non-Abelian lattice gauge theory in a ladder geometry. To exclude ETH violations due to the extensive number of local symmetries, we mapped the ladder to an effective one-dimensional model through a gauge fixing procedure.
On top of the standard Kogut--Susskind Hamiltonian for a pure gauge ladder geometry, we included a tunable deformation of the plaquette operator that breaks its invariance under group conjugation. This modification has the twofold effect of lifting the residual global $D_3\times D_3$ symmetries in the Hamiltonian spectrum and introducing a ``twisted'' interaction that is a non-Abelian extension of chiral Potts models.

Using the minimal gap ratio, we identified parameter regimes that deviate strongly from the Wigner--Dyson prediction as well as GOE-like regimes. The GOE-compatible region grows with system size, suggesting the emergence of a robust chaotic regime. In this regime, the system shows clear ETH signatures, including spectral properties of expectation values and transition matrix elements of local operators, the eigenstates' entanglement entropy, and the long-time behavior of a subsystem's entanglement Hamiltonian. 

Our study highlights how discrete non-Abelian groups are ideal toy models to test the phenomenology of far-from-equilibrium dynamics of non-Abelian LGTs, in a setting that allows for exact numerical calculations, even if limited to small sizes. Such models display rich physics, including chaotic regimes, and can provide precious intuition for more complex models relevant for fundamental physics, such as SU(N) LGTs.

Several direct avenues for future research appear laid out, though some of them imply significant computational challenges. 
For example, it would be interesting to understand to what extent our results generalize to extended two-dimensional lattices, to the presence of dynamical charges, or to other (discrete) non-Abelian gauge groups. 
Another relevant question is the role of exotic local excitations such as gluelumps, which are stable for pure $D_3$ but not $D_4$ LGTs~\cite{Popov2025}, and whether their presence or absence leads to a qualitative difference in the thermalization properties. 

\begin{acknowledgements}
The authors thank David Pascual Solis and Michele Burrello for valuable suggestions.
This project has received funding from the Swiss State Secretariat for Education, Research and Innovation (SERI) under contract number UeMO19-5.1.
This work was supported by the Provincia Autonoma di Trento and Q@TN, the joint lab between the University of Trento, FBK—Fondazione Bruno Kessler, INFN—National Institute for Nuclear Physics, and CNR—National Research Council.
M.V. acknowledges support from the Spanish Agencia Estatal de Investigaci\'on through the grant ``PID2024-161474NB-I00'', funded by MICIU/AEI/10.13039/501100011033 and co-funded by FEDER, EU.
E.B. is supported by the ERC Starting grant QS-Gauge (Grant No. 101220401). Views and opinions expressed are however those of the author(s) only and do not necessarily reflect those of the European Union or the European Research Council Executive Agency. Neither the European Union nor the granting authority can be held responsible for them.

The authors acknowledge the use of Anthropic Claude Opus 5.5 AI tools for proofreading the manuscript, verifying the correctness of the mathematical derivation, and checking possible inconsistencies between equations and numerical data. References suggested by the AI have been evaluated by the authors before being added to the manuscript.
\end{acknowledgements}

\clearpage
\onecolumngrid
\appendix

\section{Hamiltonian formulation}
\label{app:Ham}
In this Appendix, we derive the effective Hamiltonian used in the main text, see Eq.~\eqref{eq:Hfluxladder}. Starting from the ladder formulation, we fix the gauge redundancy by setting all leg links to the identity (see Sec.~\ref{app:gauge_fixing}), leaving only the rung variables as physical degrees of freedom. In the gauge-fixed basis, the electric term on the rung remains invariant (Sec.~\ref{app:rung_contribution}), while the electric terms on the legs generate non-local string operators (Sec.~\ref{app:leg_contribution}).
\begin{figure}[t]
    \centering
    \includegraphics[]{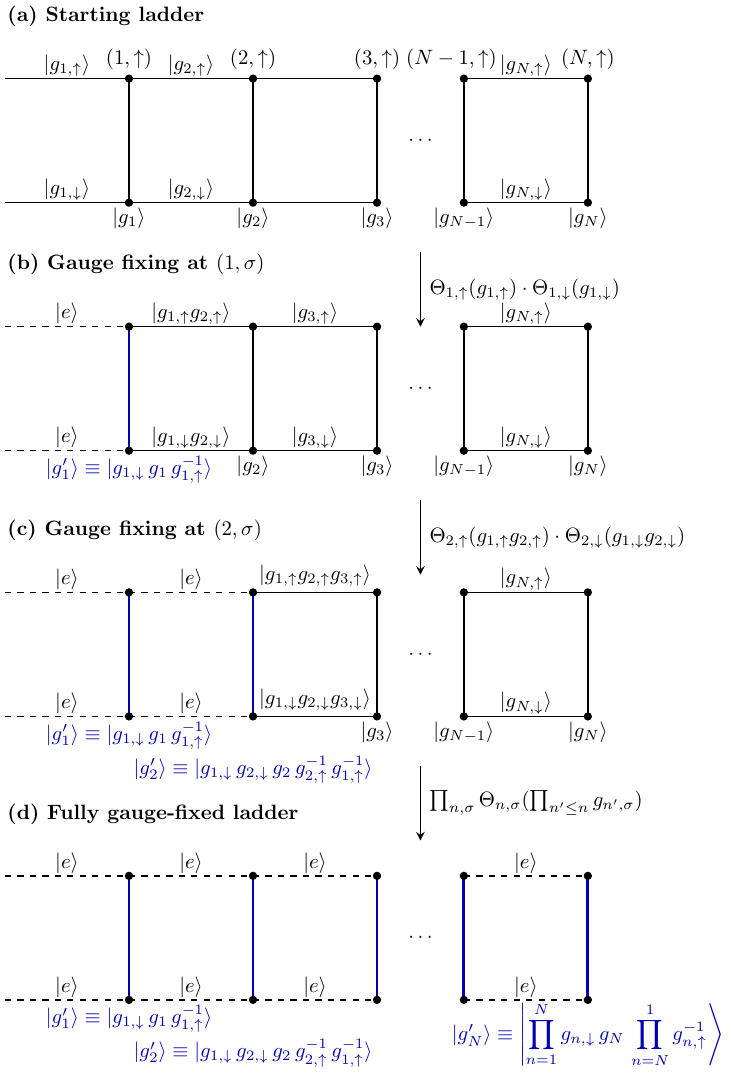}
    \caption[gauge fixing procedure]{ \justifying (a) Initial configuration of the ladder, in which each link is in a state associated with its Hilbert space labeled by a group element. (b) The ladder after the gauge fixing on the vertices $(1,\uparrow)$ and $(1,\downarrow)$. The first two legs on the left are set to the identity, and the modified rung state $\ket{g_1'}$ is redefined accordingly. (c) The ladder after the gauge fixing on the vertices $(2,\uparrow)$ and $(2,\downarrow)$. (d) Final gauge-fixed configuration of the ladder through the transformation in Eq.~\eqref{eq:gauge_fixing_trans}.}
    \label{fig:ladder}
\end{figure}
\subsection{Gauge fixing procedure}
\label{app:gauge_fixing}
Let us consider a ladder with $3N$ link variables, including the dangling legs on the left edge, $N$ of which are rungs. 
The Hilbert space can be written in the group element basis as $\mathcal{H} = \mathrm{span}\{\otimes_\ell \ket{g_\ell},\, g_\ell \in G\}$.
We denote by $\ket{\vec{g}} \equiv \bigotimes_{\ell} \ket{g_\ell}$
a basis state in the group-element basis, where \(g_\ell\in D_3\) labels the group element associated with the \(\ell\)-th link of the ladder.
Two basis states $\ket{\vec{g}}$ and $\ket{\vec{g}'}$ belong to the same gauge orbit if there exists a set of group elements $\{h_n\}$ such that $\prod_n \Theta_n(h_n)\ket{\vec{g}} = \ket{\vec{g}'}$, where $n$ labels the vertices of the ladder.
The gauge orbit of $\ket{\vec{g}}$ is therefore the set
\begin{equation}
    \mathcal{O}_{{\vec{g}}} = \left\{\prod_n \Theta_n(h_n)\ket{\vec{g}}   \quad  | \quad h_n \in G \; , \forall n \right\}.
\end{equation}
Since $[H, \Theta_n(h)] = 0$ for all $n, h$, if $\ket{\vec{g}'} = W\ket{\vec{g}}$ and $\ket{\vec{h}'} = W\ket{\vec{h}}$ for the same gauge transformation $W$, then $\bra{\vec{g}'}H\ket{\vec{h}'} = \bra{\vec{g}}H\ket{\vec{h}}$. This redundancy of information, typical of LGTs, can be reduced by fixing some of the gauge fields.
In particular, each vertex of the ladder allows for the removal of one redundant degree of freedom. In the ladder geometry with two ``dangling'' legs, such as the one depicted in Fig.~\ref{fig:ladder}(a), this means that we can fix all the leg variables and set them to the identity $\ket{e}$. 

Let us start from a state $\ket{\vec{g}}$ in the full Hilbert space. 
We denote the vertices connected to the $n-$th rung with $(n,\sigma)$, where $\sigma = \uparrow,\downarrow$ indicates whether the vertex is on the top or the bottom leg. 
We label the states on the $n$-th rung with $\ket{g_n}$ and those on the links on the left of it with $\ket{g_{n,\sigma}}$.  To proceed with the gauge fixing, we exploit the Gauss law over each vertex $(n,\sigma)$ to fix the link $\ket{g_{n,\sigma}}$ to the identity. On a ladder, the action of the gauge transformation on the three links connected to a vertex reads 
\begin{equation}
\label{eq:ladder_symm}
    \Theta_{n, \uparrow (\downarrow)} (g) = {\theta^R_{n,\uparrow(\downarrow)}(g)\, \theta^{R(L)}_n(g)\, \theta^L_{n+1,\uparrow(\downarrow)}}(g) \ .
\end{equation}
The pair of extra links on the left of the ladder plays an essential role: since Gauss's law is not imposed at their open ends, no gauge transformation other than the identity leaves a link configuration invariant. As a consequence, all gauge orbits contain the same number of elements, and the gauge-fixed description is equivalent to the gauge-invariant one~\cite{Mariani2024}.

Specifically, the gauge fixing procedure goes as follows.
Starting from the state of the first rung and the two left attached links $\ket{g_{1,\downarrow},g_1,g_{1,\uparrow}}$, we apply a gauge transformation at the vertices $(1,\sigma)$, associated with the element $g_{1,\sigma}$, leading to 
\begin{equation}
    \Theta_{1,\uparrow}(g_{1,\uparrow})\Theta_{1,\downarrow}(g_{1,\downarrow}) \ket{g_{1,\downarrow},g_1,g_{1,\uparrow}} = \ket{e, g_{1,\downarrow}g_1g_{1,\uparrow}^{-1},e} \ .
\end{equation}
Note that the rung state is modified according to the different orientation of the rung with respect to the two vertices.
In addition, from Eq.~\eqref{eq:ladder_symm} we see that the second-leftmost legs are also changed:
\begin{equation}
    \Theta_{1,\uparrow}(g_{1,\uparrow})\Theta_{1,\downarrow}(g_{1,\downarrow}) \ket{g_{2,\downarrow},g_2,g_{2,\uparrow}} = \ket{g_{1,\downarrow}g_{2,\downarrow},g_2,g_{1,\uparrow}g_{2,\uparrow}} \ .
\end{equation}
This first step of the gauge fixing procedure is depicted in Fig.~\ref{fig:ladder}(b).

In the next step of the gauge fixing, the transformation will be with respect to the product $\ket{g_{1,\sigma}g_{2,\sigma}}$, as shown in Fig.~\ref{fig:ladder}(c), to set the second pair of legs to the identity. The process then proceeds iteratively, and the string of group products will spread on all the rungs up to the end of the ladder, as shown in Fig.~\ref{fig:ladder}(d).
The total gauge fixing procedure can be written as
\begin{equation}\label{eq:gauge_fixing_trans}
    \ket{\vec{g}} \rightarrow \ket{\vec{g}_{{\rm g.f.}}} = \prod_{n,\sigma} \Theta_{n,\sigma}\left( \prod_{n'\leq n} g_{n',\sigma} \right) \ket{\vec{g}} \ .
\end{equation}
As a result, we get an effective basis that has non-trivial components only on the rungs, as shown in Ref.~\cite{Munk2018}, 
\begin{align}\label{eq:gauge_fixing}
    & \ket{g_n} \rightarrow \ket{\prod_{n'=1}^n g_{n',\downarrow} g_n \prod_{n'=n}^1 g_{n',\uparrow}^{-1}}\equiv \ket{g'_n}=\ket{g_{{\rm g.f.},n}} \ , \nonumber \\
    & \ket{g_{n,\sigma}} \rightarrow \ket{e} \ .
\end{align} 

It is now possible to rewrite the action of the magnetic and electric terms of the Hamiltonian in the new gauge-fixed basis, given by Eq.~\eqref{eq:gauge_fixing}. Each plaquette product in the magnetic term simplifies from a $4-$body term to a $2-$ body term. Since the two links on the legs are fixed to the identity, the non-trivial product is given by the two rungs, 
\begin{equation}\label{eq:h_b_gauge_fix}
    H_B = - \frac{1}{\mathrm{g}^2}\sum_p \Re\left(\textrm{Tr}\left[U_{p1}U_{p2}U_{p3}^\dagger U_{p4}^\dagger\right] \right) \rightarrow H_B^{\textrm{gf}} = - \frac{1}{\mathrm{g}^2}\sum_{r=1}^{N-1} \Re \left(\textrm{Tr}\left[U_r^\dagger U_{r+1}\right] \right) \ ,\end{equation}
where the sum is over all the rungs $r$.
Regarding the electric term, we have to take into account contributions on the rungs and on the legs. For convenience, we derive these in two distinct subsections.

\subsection{Rung contribution}
\label{app:rung_contribution}
In the following, we examine the effect of the gauge fixing transformation on the matrix elements of $H_E$, the electric Hamiltonian. The full gauge fixing is implemented by the unitary transformation
\begin{equation}
    V_g = \prod_{n,\sigma} \Theta_{n,\sigma}\!\left(\prod_{n' \leq n} g_{n',\sigma}\right) .
\end{equation} 
The matrix element of $H_E$ in the gauge-fixed basis becomes
\begin{equation}
    \langle {\vec{g}} | H_E | {\vec{h}} \rangle 
    \;\rightarrow\; 
    \langle {\vec{g}}| V_g^\dagger V_g   H_E V_h^\dagger V_h | {\vec{h}} \rangle =\langle {\vec{g}_r} | V_g H_E\, V_h^\dagger | {\vec{h}_r} \rangle ,
\end{equation}
where $\ket{\vec{g}_r}\equiv V_g\ket{\vec{g}}$ denotes the gauge-fixed basis state, with non-trivial group elements on the rungs only.
Since $H_E = \bigoplus_\ell H_E^{(\ell)}$ is a direct sum of single-link terms, we focus on the term acting on rung $w$, which connects the vertices $(w,\uparrow)$ and $(w,\downarrow)$, denoted $H_{E,w}$. Defining the string
\begin{equation}
    \tilde{g}_{n,\sigma} \equiv \prod_{n' \leq n} g_{n',\sigma} ,
\end{equation}
the action of $H_{E,w}$ on the gauge-fixed basis is given by
\begin{equation}
    H_{E,w} \;\longrightarrow\; V_g H_{E,w}\, V_h^\dagger =
    \prod_{n,\sigma} \Theta_{n,\sigma}(\tilde{g}_{n,\sigma})\; 
    H_{E,w} \;
    \prod_{n',\sigma} \Theta_{n',\sigma}(\tilde{h}^{-1}_{n',\sigma}) \ .
\end{equation}
Since $H_{E,w}$ acts locally on the rung $w$ only, the factors $\Theta_{n,\sigma}$  with $n,\sigma \neq w,\sigma$ commute with $H_{E,w}$, and the product $\Theta_{n,\sigma}(\tilde{g}_{n,\sigma})\Theta_{n',\sigma}(\tilde{h}^{-1}_{n',\sigma})$ gives $\delta_{\tilde{g}_{n,\sigma},\tilde{h}_{n',\sigma}}$ for all $\ell \neq w$. Indeed, in the matrix element between gauge-fixed states, the condition that these factors yield a non-zero contribution forces $\tilde{g}_{n,\sigma} = \tilde{h}_{n,\sigma}$ for all $n,\sigma \neq w,\sigma$. We are therefore left with
\begin{align}
    \langle {\vec{g}_r} | V_g & H_{E,w}\, V_h^\dagger | {\vec{h}_r} \rangle = \nonumber \\
    &\bra{\vec{g}_r} \left( \prod_{\ell\neq w,(\sigma)} \delta_{\tilde{g}_{n,\sigma},\tilde{h}_{n',\sigma}} \right)  \Theta_{w,\uparrow}(\tilde{g}_{w,\uparrow})\, \Theta_{w,\downarrow}(\tilde{g}_{w,\downarrow})\, H_{E,w}^{(1)}\, \Theta_{w,\uparrow}(\tilde{h}^{-1}_{w,\uparrow})\, \Theta_{w,\downarrow}(\tilde{h}^{-1}_{w,\downarrow}) \ket{\vec{h}_r} \,.
\end{align}
Using the expression for the single-link electric Hamiltonian,
\begin{equation}
    H_{E,\ell}^{(1)} = \mathrm{g}^2\sum_j \frac{\dim(j)}{|G|}\, \alpha_j \sum_g \chi^j(g)\, \theta_\ell^L(g) \ ,
\end{equation}
we can write explicitly the matrix element on the links variable where the action is not trivial per se, i.e., on the rung $w$ and the $4$ links connected to it:
\begin{align}
    \langle {\vec{g}_r} | V_g H_{E,w}\, V_h^\dagger | {\vec{h}_r} \rangle =& \ \mathrm{g}^2\sum_j \frac{\dim(j)}{|G|}\, \alpha_j \sum_g \chi^j(g) 
    \nonumber \\
    &\prod_\sigma\bra{e_{w, \sigma}}  \theta^R_{w, \sigma} (\tilde{g}_{w, \sigma}) \theta^R_{w, \sigma} (\tilde{h}^{-1}_{w, \sigma}) \ket{e_{w, \sigma}} \nonumber \\
    &\prod_\sigma\bra{e_{w+1,\sigma}}  \theta^L_{w+1,\sigma} (\tilde{g}_{w,\sigma}) \theta^L_{w+1,\sigma} (\tilde{h}^{-1}_{w,\sigma}) \ket{e_{w+1,\sigma}} \nonumber \\
    &\bra{g_w} \theta^L_w(\tilde{g}_{w,\downarrow})\theta^L_w(g)\theta^L_w(\tilde{h}^{-1}_{w,\downarrow}) \theta^R_w(\tilde{g}_{w,\uparrow})\theta^R_w(\tilde{h}^{-1}_{w,\uparrow}) \ket{h_w} \ .
\end{align}
The matrix elements over the four legs give a non-zero contribution only when $\tilde{g}_{w,\sigma}=\tilde{h}_{w,\sigma}$. This condition gives a contribution $\delta_{\tilde{g}_{w,\sigma},\tilde{h}_{w,\sigma}}$, which we can use to simplify the fourth row of the equation. So, the matrix element in the gauge-fixed basis is limited to the action over the rung $w$, where the product $\delta_{\tilde{g}_{w,\uparrow},\tilde{h}_{w,\uparrow}} \theta^R_w(\tilde{g}_{w,\uparrow})\theta^R_w(\tilde{h}^{-1}_{w,\uparrow}) = \mathbb{I}$, and we get
\begin{equation}
    \langle {\vec{g}_r} | V_g H_{E,w}\, V_h^\dagger | {\vec{h}_r} \rangle = \mathrm{g}^2\sum_j \frac{\dim(j)}{|G|}\, \alpha_j \sum_g \chi^j(g) \bra{g_w} \theta^L_w(\tilde{g}_{w,\downarrow})\theta^L_w(g)\theta^L_w(\tilde{g}^{-1}_{w,\downarrow})\ket{h_w} \ .
\end{equation}
Since $\theta^L_w(\tilde{g}_{w,\downarrow})\theta^L_w(g)\theta^L_w(\tilde{g}^{-1}_{w,\downarrow}) = \theta_w^L(\tilde{g}_{w,\downarrow}\, g\, \tilde{g}_{w,\downarrow}^{-1})$ is the conjugate of $g$, we obtain  $\chi^j\!\left(\tilde{g}_{w,\downarrow}\, g\, \tilde{g}_{w,\downarrow}^{-1}\right) = \chi^j(g)$, because characters are class functions. The matrix element of $H_{E,w}$ in the gauge-fixed basis therefore yields the same contribution as the original rung electric term, meaning that the form of the rung electric Hamiltonian is unchanged by the gauge fixing procedure:
\begin{equation}
   \bra{\vec{g}_r} V_g H_{E,w}\, V_h^\dagger \ket{\vec{h}_r} = \mathrm{g}^2\sum_j \frac{\dim(j)}{|G|}\, \alpha_j \sum_g \chi^j(g)\bra{g_w}\theta^L_w(g)\ket{h_w} \ .
\end{equation}

\subsection{Leg contribution}
\label{app:leg_contribution}
In this subsection, we consider the contribution to the gauge-fixed electric Hamiltonian from a leg link. We focus on the link connecting vertices $(w-1,\sigma)$ and $(w,\sigma)$, which we denote $\ell_{w,\sigma}$. Following the same structure as in the rung case, the gauge-fixed matrix element is
\begin{align}
    \bra{\vec{g}_r} V_g\, H_{E,w,\sigma}\, V_h^\dagger \ket{\vec{h}_r} = \bra{\vec{g}_r} \left(\prod_{n,\sigma} \Theta_{n,\sigma}(\tilde{g}_{n,\sigma})\right) H_{E,w,\sigma} \left(\prod_{n',\sigma} \Theta_{n',\sigma}(\tilde{h}^{-1}_{n',\sigma})\right) \ket{\vec{h}_r} ,
\end{align}
where all the leg links are gauge-fixed to $\ket{e}$. For all links at positions $n < w$, the matrix element $\bra{\vec{g}_r} V_g\, H_{E,w,\sigma}\, V_h^\dagger \ket{\vec{h}_r}$ gives $\delta_{\tilde{g}_{n,\sigma},\tilde{h}_{n,\sigma}}$ as in the rung case, constraining $\tilde{g}_{n,\sigma} = \tilde{h}_{n,\sigma}$ for $n < w$. We are left with the action on $\ell_{w,\sigma}$ and on links at positions $n \ge w$.

We also write for convenience $H_{E,w,\sigma}^{(1)} = \mathrm{g}^2\sum_j\frac{\dim(j)}{|G|}\alpha_j \sum_g \chi^j(g)\,\theta^L_{w,\sigma}(g)$. In the following, we fix $\sigma = \downarrow$ for simplicity, but the equivalent procedure also works for $\sigma = \uparrow$. The consequence is that we get $\delta_{\tilde{g}_{n,\uparrow},\tilde{h}_{n,\uparrow}}$ also on the links on the top leg, 

\begin{align}
    \bra{\vec{g}_r} V_g&\, H_{E,{w,\downarrow}}\, V_h^\dagger \ket{\vec{h}_r} = \mathrm{g}^2\sum_j \frac{\dim(j)}{|G|}\alpha_j \sum_g \chi^j(g)  \\ 
    &\bra{\vec{g}_r} \prod_{n,n'< w} \delta_{\tilde{g}_{n,(\sigma)},\tilde{h}_{n',(\sigma)}} \prod_{n,n'\ge w} \delta_{\tilde{g}_{n,\uparrow},\tilde{h}_{n',\uparrow}} 
    \Theta_{n,\downarrow}(\tilde{g}_{n,\downarrow})\theta^L_{{w,\downarrow}}(g) \Theta_{n',\downarrow}(\tilde{h}^{-1}_{n',\downarrow})
    \ket{\vec{h}_r} \ . \nonumber
\end{align}

Now we focus on the product on the links $w$ and $\ell_{w,\downarrow}$, where the non-trivial operators are $\theta^L_{{w,\downarrow}}$. We make explicit the first two terms of the product, when $n=n'=w$, such that we get
\begin{align}
    \prod_{n,n'\ge w} & \Theta_{n,\downarrow}(\tilde{g}_{n,\downarrow})\theta^L_{{w,\downarrow}}(g) \Theta_{n',\downarrow}(\tilde{h}^{-1}_{n',\downarrow}) = \theta^R_{{w,\downarrow}}(\tilde{g}_{w,\downarrow})\theta^L_{{w,\downarrow}}(g)\theta^R_{{w,\downarrow}}(\tilde{h}^{-1}_{w,\downarrow}) 
    \theta^L_{{w}}(\tilde{g}_{w,\downarrow}) \theta^L_{{w}}(\tilde{h}^{-1}_{w,\downarrow})  \nonumber \\
    &
    \prod_{n,n'\ge w+1} \Theta_{n,\downarrow}(\tilde{g}_{n,\downarrow}) \theta^L_{{w+1,\downarrow}}(\tilde{g}_{w,\downarrow}) \theta^L_{{w+1,\downarrow}}(\tilde{h}^{-1}_{w,\downarrow}) \Theta_{n',\downarrow}(\tilde{h}^{-1}_{n',\downarrow}) \ .
\end{align}
On the link $\ell_{w,\downarrow}$, we get the matrix element $\bra{e} \theta^R_{w,\downarrow}(\tilde{g}_{w,\downarrow}\tilde{h}^{-1}_{w,\downarrow}) \theta^L_{w,\downarrow}(g) \ket{e}$, which only gives non-zero contribution for $g=\tilde{f}_{w,\downarrow}$, where $\tilde{f}_{w,\downarrow}=\tilde{g}_{w,\downarrow}\tilde{h}^{-1}_{w,\downarrow}$.

Hence,
\begin{align}
    \prod_{n,n'\ge w} &\Theta_{n,\downarrow}(\tilde{g}_{n,\downarrow})\theta^L_{{w,\downarrow}}(g) \Theta_{n',\downarrow}(\tilde{h}^{-1}_{n',\downarrow}) = \nonumber \\
    &=\delta_{\tilde{f}_{w,\downarrow},g}
    \theta^L_{{w}}(\tilde{f}_{w,\downarrow}) 
    \prod_{n,n'\ge w+1} \Theta_{n,\downarrow}(\tilde{g}_{n,\downarrow}) \theta^L_{{w+1,\downarrow}}(\tilde{f}_{w,\downarrow}) \Theta_{n',\downarrow}(\tilde{h}^{-1}_{n',\downarrow})= \nonumber \\ 
    &= \theta^L_{{w}}(g) 
    \prod_{n,n'\ge w+1} \Theta_{n,\downarrow}(\tilde{g}_{n,\downarrow}) \theta^L_{{w+1,\downarrow}}(g) \Theta_{n',\downarrow}(\tilde{h}^{-1}_{n',\downarrow}) \ .
\end{align}
So, the gauge fixing procedure forces the electric field localized on the leg $\ell_{w,\sigma}$ to spread over the rung on the right and the leg $\ell_{w+1,\sigma}$.

Now, we can repeat the same procedure on the link $\ell_{w+1,\sigma}$, and so on. The final result is that the operator on the link $\ell_{w,\sigma}$ spreads over all the rungs $r\ge w$,

\begin{equation}
    \bra{\vec{g}_r} V_g\,H_{E,w,\downarrow}\,V_h^\dagger \ket{\vec{h}_r} = \mathrm{g}^2\sum_j\frac{\dim(j)}{|G|}\,\alpha_j\sum_g\chi^j(g) \prod_{r\ge w}\theta^L_{r}(g) \ .
    \label{eq:leg_string}
\end{equation}
Summing over all leg positions $w$ on both legs, and noting that on the upper leg we will get a string of $\theta^R$ operators because of the directionality, the total leg electric Hamiltonian in the gauge-fixed basis is
\begin{equation}
    H_E^{\mathrm{leg}} = \mathrm{g}^2\sum_r\sum_j\frac{\dim(j)}{|G|}\,\alpha_j\sum_g\chi^j(g)     \left[\prod_{r'\ge r}\theta^L_{r'}(g)+\prod_{r'\ge r}\theta^R_{r'}(g)\right] \ ,
\end{equation}

Hence, while the encoding of the rung terms remains the same as the original model, the electric Hamiltonian on the leg terms is responsible for a series of non-local string terms appearing in the gauge-fixed Hamiltonian.

These strings extend to the right of the rung, $r \ge w$, because the dangling links sit on the left edge of the ladder. The Hamiltonian of Eq.~\eqref{eq:Hfluxladder} used in the main text corresponds to the spatial reflection of the ladder, $r \to N+1-r$, which moves the dangling links to the right edge and turns the strings $\prod_{r'\ge r}\theta^{L/R}_{r'}(g)$ into $\prod_{r'=1}^{r}\theta^{L/R}_{r'}(g)$. All other terms of the gauge-fixed Hamiltonian are invariant under this reflection. The deformation of Eq.~\eqref{eq:Hmagmod} is then applied to the rung variables of the reflected ladder.

\newpage
\section{Additional results regarding ETH}\label{app:more_res}
In this Appendix, we show additional results supporting ETH in the system under investigation. First, we analyze the diagonal matrix elements of a local observable in the regime deviating from the Wigner--Dyson prediction, complementing the analysis done in the main text for the GOE-like region (Sec.~\ref{app:diagelempoi}). Second, we highlight the differences in the off-diagonal signatures between the chaotic regime, where ETH fully holds, and the region where at finite size ETH probes are less conclusive (Sec.~\ref{app:offdiagelem}). Finally, we repeat the ETH diagnostics of the main text for a plaquette at the boundary of the ladder (Sec.~\ref{app:boundary_plaquette}).

\subsection{Diagonal elements of local observables in the sub-GOE regime}
\label{app:diagelempoi}
\begin{figure*}[!htb]
    \centering
    \includegraphics[width=\textwidth]{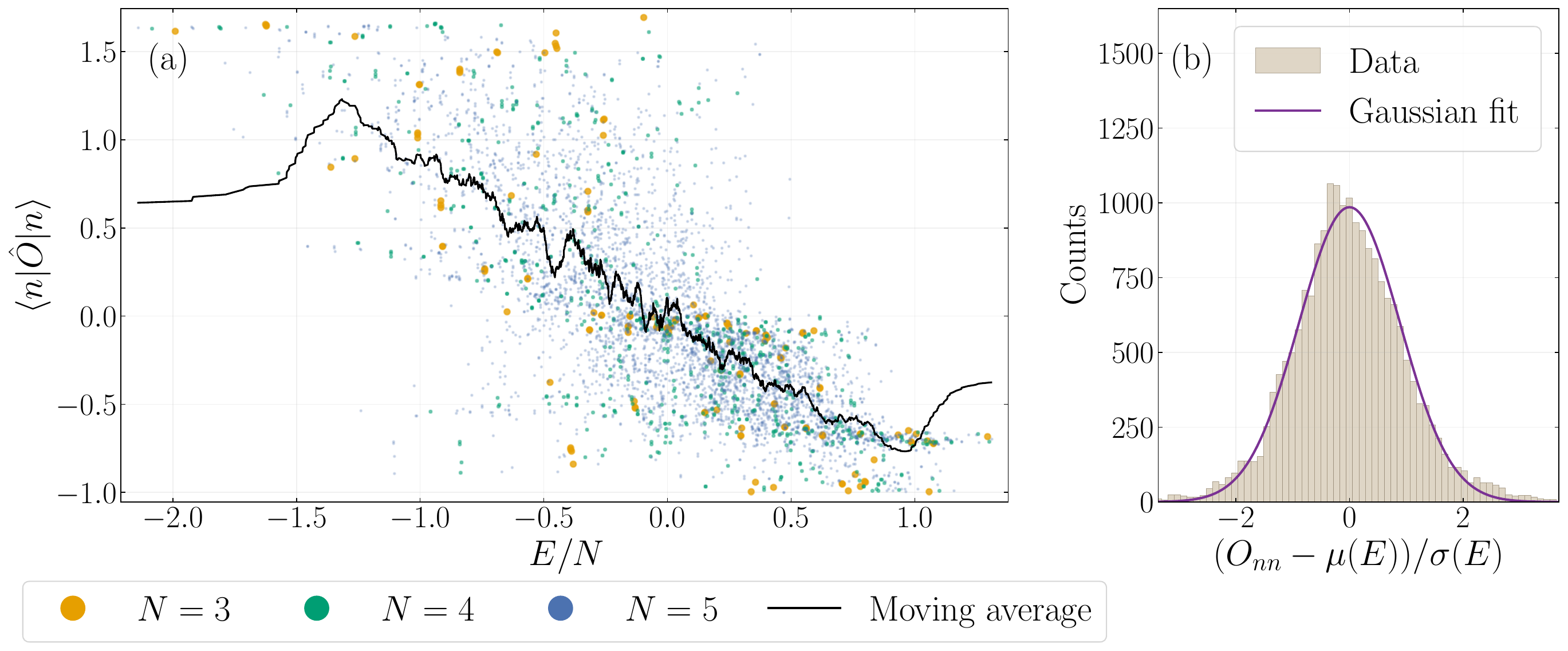}
    \caption{ \justifying
    Diagonal matrix elements $O_{nn}=\langle n|\hat{O}|n\rangle$ of the single-plaquette operator as a function of the energy density $E/N$. Panel~(a) shows a representative point in parameter space, $\mathrm{g}=0.68$ and $\varepsilon=0.06$.
    The black curve denotes the moving average computed from the $N=5$ data.
    For the limited system sizes we were able to tackle numerically, the diagonal matrix elements do not clearly approach a smooth function of the energy, in contrast to what was observed in Fig.~\ref{fig:diagonal_eth}.
    Panel~(b) shows the normalized residuals $(O_{nn}-\mu(E))/\sigma(E)$ obtained from the $N=5$ data in panel (a), together with a Gaussian fit, which matches the numerics well. The data are obtained considering randomly chosen energy windows with $\Delta E = 0.2$ in the bulk of the spectrum.}
    \label{fig:diagonal_eth_app}
\end{figure*}

First, we study the behavior of the diagonal elements of local observables in the region of the phase diagram where the level-spacing statistics deviates significantly from GOE, to be compared with the analogous results deep within the GOE region, Fig.~\ref{fig:diagonal_eth}.
We repeat the same analysis for different system sizes $N=3,4,5$, considering a parameter couple $({\rm g}, \varepsilon)$ where $\braket{r}<\braket{r}_{\rm GOE}$. Figure~\ref{fig:diagonal_eth_app}(a) reports our numerical results, showing a partial violation of ETH.
In this regime, the diagonal matrix elements do not clearly organize into a smooth function of energy. Moreover, their fluctuations do not show an evident decrease with increasing system size. 
Since the system sizes we were able to approach numerically are rather limited, and judging from the growth of the GOE-compatible parameter region, we expect this behavior to be a finite-size effect with full agreement with ETH emerging for sufficiently large $N$. 
As an observation in favor of this expectation, the fluctuations of the diagonal elements around an energy-binned average are well described by a Gaussian distribution, see Fig.~\ref{fig:diagonal_eth_app}(b), as expected from random matrix theory and ETH.

\subsection{Off-diagonal elements of local observables}
\label{app:offdiagelem}
Next, we characterize the off-diagonal matrix elements of the plaquette operator $\hat{O}$ through the ETH lens.
Recalling Eq.~\eqref{eq:ETH_general}, they should follow
\begin{equation}
    O_{mn} = e^{-S(\overline{E})/2}f_O(\overline{E},\omega)R_{mn} \ ,
    \label{eq:offdiagETH}
\end{equation}
where $\overline{E}=(E_m+E_n)/2$ and $\omega=E_n-E_m$.
We first compare the off-diagonal matrix elements at the two representative points of the parameter space considered above.
\begin{figure*}[!htb]
\centering
\includegraphics[width=\textwidth]{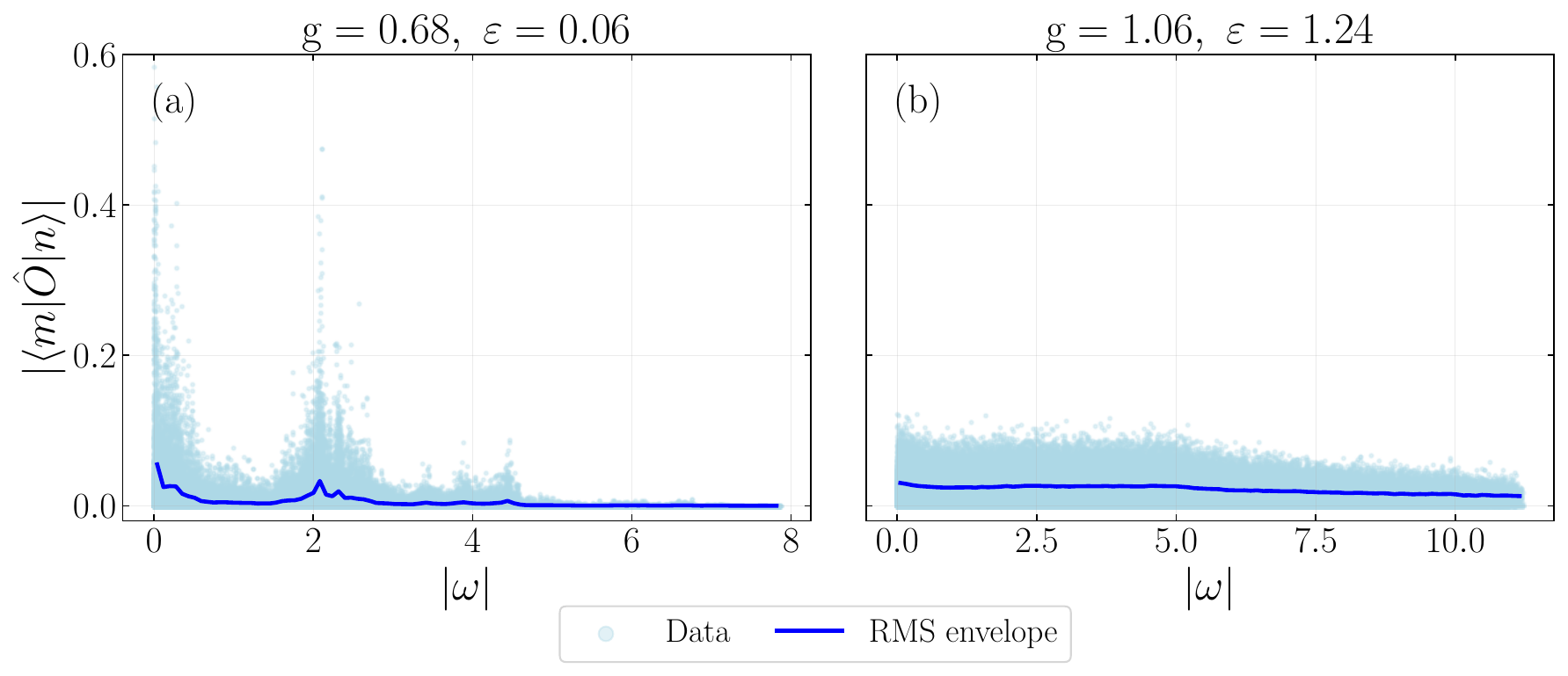}
\caption{\justifying
    Absolute value of the off-diagonal matrix elements $|O_{mn}|$ of the single-plaquette operator $\hat{O}$ as a function of the energy difference $\omega=|E_m-E_n|$, for $N=5$. The analysis is restricted to eigenstates in the bulk of the spectrum. Light-blue points denote the individual off-diagonal matrix elements, while the dark-blue curve shows the RMS envelope obtained by binning the data in $\omega$. 
    In the regime where the spectral statistics deviate from GOE, strong fluctuations are observed (left), while results in the regime compatible with GOE (right) are significantly smoother. }
\label{fig:offdiagonal_eth_app}
\end{figure*}

In Fig.~\ref{fig:offdiagonal_eth_app}, we compare the off-diagonal matrix elements of the single-plaquette operator in a regime where the spectral statistics (a) deviate from and (b) agree with the GOE prediction. To characterize their frequency dependence, we group the matrix elements into bins of $\omega=|E_m-E_n|$ and compute the RMS envelope $\overline{|O_{mn}|}_{\mathrm{RMS}}(\omega)$ in each bin. This provides an estimate of the typical magnitude of the off-diagonal matrix elements as a function of frequency and is directly related to the smooth ETH function $f_O(\overline{E},\omega)$.

The two regimes display qualitatively different behavior. In the first case, the off-diagonal matrix elements show strong irregularities, including a large number of very small or vanishing values together with isolated large outliers. This indicates that, for the observable considered here, the off-diagonal matrix elements are not described by a smooth ETH-like profile. In contrast, in the GOE-compatible regime the data form a more regular frequency-dependent structure around the RMS envelope, without the same proliferation of zeros and large outliers. 

\subsection{ETH signatures for boundary plaquette observable}
\label{app:boundary_plaquette}

\begin{figure*}[!htb]
    \centering
    \includegraphics[width=\linewidth]{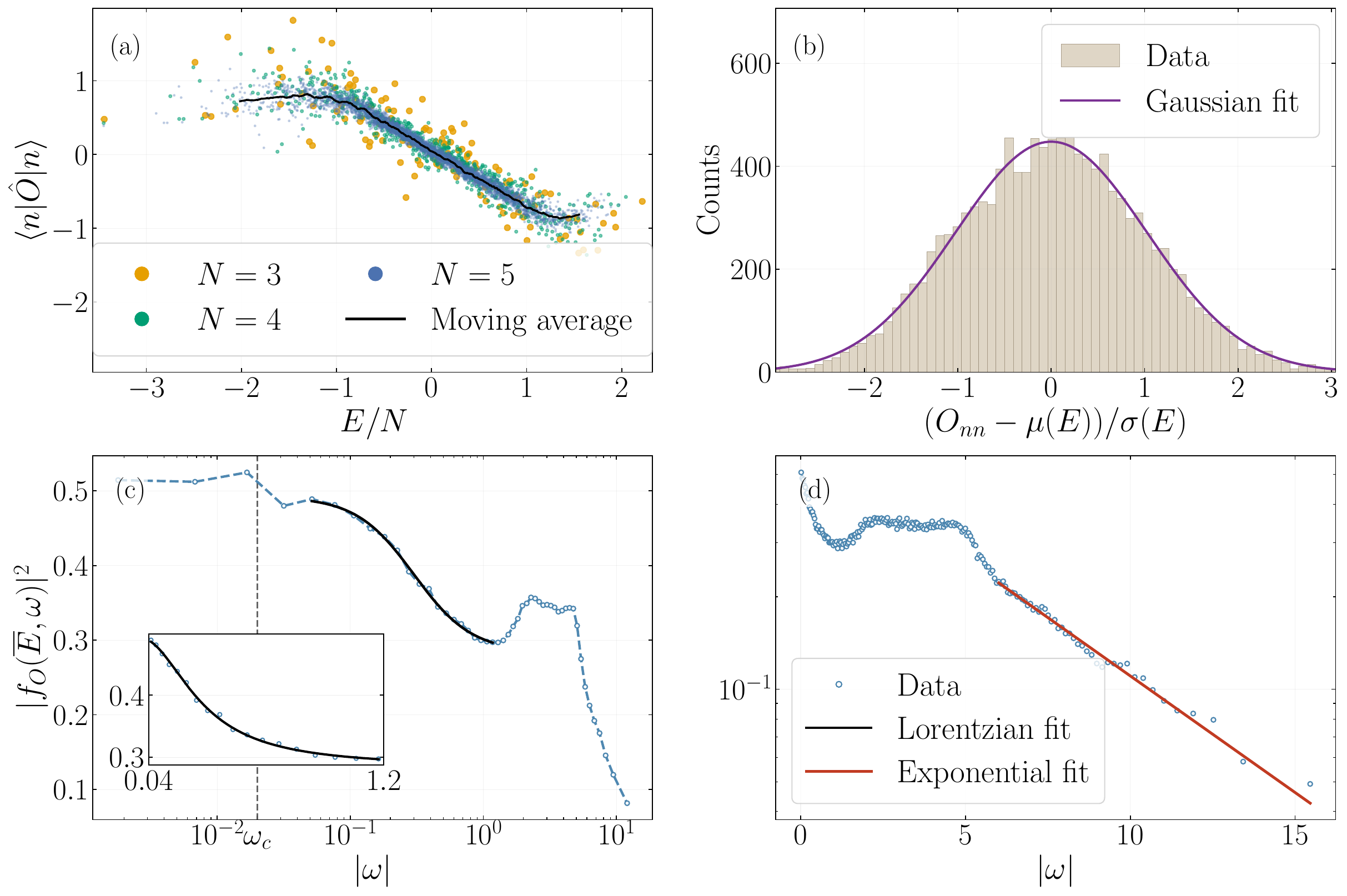}
    \caption{\justifying
    ETH diagnostics for the single-plaquette operator $\hat{O}$ acting on the leftmost plaquette of the ladder, at the representative point with GOE level statistics $\mathrm{g}=1.06$ and $\varepsilon=1.24$. (a) Diagonal matrix elements $O_{nn}=\langle n|\hat{O}|n\rangle$
    as a function of the energy density $E/N$. The black curve denotes the running average for $N=5$. (b) Normalized residuals $(O_{nn}-\mu(E))/\sigma(E)$ for $N=5$, together with a Gaussian fit (purple solid line). (c-d) Frequency dependence of the squared spectral function
    $|f_O(\overline{E},\omega)|^2$ extracted from the off-diagonal matrix elements for $N=5$.
    Panel (c) highlights the low-frequency behavior and the Lorentzian fit, while panel (d) shows the high-frequency exponential tail. The parameters and numerical procedures are the same as
    in Fig.~\ref{fig:diagonal_eth}, where the central plaquette is considered.
    }
    \label{fig:boundary_plaquette}
\end{figure*}

For completeness, we report the ETH diagnostics obtained using the single-plaquette operator acting on the leftmost plaquette of the ladder, rather than
the central plaquette considered in the main text.
Figure~\ref{fig:boundary_plaquette} shows the corresponding results for the same parameters and numerical procedures used in Fig.~\ref{fig:diagonal_eth}.

The main qualitative difference appears in the squared spectral function $|f_O(\overline{E},\omega)|^2$, where the intermediate-frequency feature, appearing as a knee for the central plaquette, develops into a broad peak for the boundary plaquette.

\newpage
\bibliography{biblio}

\end{document}